\PassOptionsToPackage{unicode}{hyperref}
\PassOptionsToPackage{hyphens}{url}
\documentclass[hidelinks]{article}
\usepackage[margin=1in]{geometry}
\usepackage{amsmath,amssymb}
\usepackage{kpfonts}
\usepackage{iftex}
\ifPDFTeX
  \usepackage[T1]{fontenc}
  \usepackage[utf8]{inputenc}
  \usepackage{textcomp} 
\else 
  \usepackage{unicode-math}
  \defaultfontfeatures{Scale=MatchLowercase}
  \defaultfontfeatures[\rmfamily]{Ligatures=TeX,Scale=1}
\fi
\IfFileExists{upquote.sty}{\usepackage{upquote}}{}
\IfFileExists{microtype.sty}{
  \usepackage[]{microtype}
  \UseMicrotypeSet[protrusion]{basicmath} 
}{}
\makeatletter
\@ifundefined{KOMAClassName}{
  \IfFileExists{parskip.sty}{%
    \usepackage{parskip}
  }{
    \setlength{\parindent}{0pt}
    \setlength{\parskip}{6pt plus 2pt minus 1pt}}
}{
  \KOMAoptions{parskip=half}}
\makeatother
\usepackage{xcolor}
\IfFileExists{xurl.sty}{\usepackage{xurl}}{} 
\IfFileExists{bookmark.sty}{\usepackage{bookmark}}{\usepackage{hyperref}}
\usepackage{color}
\usepackage{fancyvrb}

\DefineVerbatimEnvironment{Highlighting}{Verbatim}{commandchars=\\\{\}}
\usepackage{framed}
\definecolor{shadecolor}{RGB}{248,248,248}
\newenvironment{Shaded}{\begin{snugshade}}{\end{snugshade}}

\newcommand{\AttributeTok}[1]{\textcolor[rgb]{0.23,0.49,0.23}{#1}}

\newcommand{\BuiltInTok}[1]{#1}
\newcommand{\CharTok}[1]{\textcolor[rgb]{0.31,0.60,0.02}{#1}}
\newcommand{\CommentTok}[1]{\textcolor[rgb]{0.56,0.35,0.01}{#1}}

\newcommand{\ControlFlowTok}[1]{\textcolor[rgb]{0.29,0.53,0.83}{\textbf{#1}}}
\newcommand{\DataTypeTok}[1]{\textcolor[rgb]{0.13,0.29,0.53}{#1}}
\newcommand{\DecValTok}[1]{\textcolor[rgb]{0.00,0.00,0.81}{#1}}

\newcommand{\ImportTok}[1]{#1}

\newcommand{\KeywordTok}[1]{\textcolor[rgb]{0.13,0.29,0.53}{\textbf{#1}}}
\newcommand{\NormalTok}[1]{#1}
\newcommand{\OperatorTok}[1]{\textcolor[rgb]{0.81,0.36,0.00}{\textbf{#1}}}

\newcommand{\PreprocessorTok}[1]{\textcolor[rgb]{0.56,0.35,0.01}{\textit{#1}}}

\newcommand{\SpecialCharTok}[1]{\textcolor[rgb]{0.70,0.70,0.00}{#1}}

\newcommand{\StringTok}[1]{\textcolor[rgb]{0.31,0.60,0.02}{#1}}

\providecommand{\tightlist}{%
  \setlength{\itemsep}{0pt}\setlength{\parskip}{0pt}}
\ifLuaTeX
  \usepackage{selnolig}  
\fi
\usepackage{fontawesome5}

\begin{document}

\title{A C++17 Thread Pool for High-Performance Scientific Computing}

\author{
  \textbf{Barak Shoshany}\\
  \\
  \small \faIcon{github} \href{https://github.com/bshoshany}{@bshoshany}\\
  \small \faIcon{envelope} \href{mailto:bshoshany@brocku.ca}{bshoshany@brocku.ca}\\
  \small \faIcon{globe} \href{https://baraksh.com/}{https://baraksh.com/}\\
  \small \faIcon{orcid} \href{https://orcid.org/0000-0003-2222-127X}{0000-0003-2222-127X}\\
  \small \faIcon{university} \href{https://brocku.ca/}{Department of Physics, Brock University}\\
  \small \faIcon{map-marker-alt} \href{https://goo.gl/maps/qscBMigohESxxczM7}{1812 Sir Isaac Brock Way, St. Catharines, Ontario, L2S 3A1, Canada}
}

\date{}

\maketitle

\renewenvironment{abstract}
{\quotation\small\noindent\rule{\linewidth}{.5pt}\par\smallskip
  {\centering\bfseries\abstractname\par}\medskip}
{\par\noindent\rule{\linewidth}{.5pt}\endquotation}

\begin{abstract}
  We present a modern C++17-compatible thread pool implementation, built
  from scratch with high-performance scientific computing in mind. The thread pool is
  implemented as a single lightweight and self-contained class, and does
  not have any dependencies other than the C++17 standard library, thus
  allowing a great degree of portability. In particular, our
  implementation does not utilize OpenMP or any other high-level
  multithreading APIs, and thus gives the programmer precise low-level
  control over the details of the parallelization, which permits more robust
  optimizations. The thread pool was extensively tested on both AMD and Intel CPUs
  with up to 40 cores and 80 threads. This paper provides motivation, detailed usage
  instructions, and performance tests.
\end{abstract}

\tableofcontents

The source code for this package is freely available for download on GitHub:

\begin{center}
  \url{https://github.com/bshoshany/thread-pool}
\end{center}

This companion paper on arXiv is updated infrequently, and may be out of date. It is currently updated up to v3.0.0 (2022{-}05{-}30) of the library. The \texttt{README.md} file on \href{https://github.com/bshoshany/thread-pool}{the GitHub repository} is guaranteed to always be up to date.

\hypertarget{introduction}{%
  \section{Introduction}\label{introduction}}

\hypertarget{motivation}{%
  \subsection{Motivation}\label{motivation}}

Multithreading \cite{Williams2019} is essential for modern high-performance computing. Since
C++11, the C++ \cite{Stroustrup2013}\cite{Stroustrup2014}\cite{Stroustrup2018} standard library has included built-in low-level
multithreading support using constructs such as \texttt{std::thread}.
However, \texttt{std::thread} creates a new thread each time it is
called, which can have a significant performance overhead. Furthermore,
it is possible to create more threads than the hardware can handle
simultaneously, potentially resulting in a substantial slowdown.

The library presented here contains a thread pool class,
\texttt{BS::thread\_pool}, which avoids these issues by creating a fixed
pool of threads once and for all, and then continuously reusing the same
threads to perform different tasks throughout the lifetime of the
program. By default, the number of threads in the pool is equal to the
maximum number of threads that the hardware can run in parallel.

The user submits tasks to be executed into a queue. Whenever a thread
becomes available, it retrieves the next task from the queue and
executes it. The pool automatically produces an \texttt{std::future} for
each task, which allows the user to wait for the task to finish
executing and/or obtain its eventual return value, if applicable.
Threads and tasks are autonomously managed by the pool in the
background, without requiring any input from the user aside from
submitting the desired tasks.

The design of this package was guided by four important principles.
First, \emph{compactness}: the entire library consists of just one small
self-contained header file, with no other components or dependencies.
Second, \emph{portability}: the package only utilizes the C++17 standard
library \cite{ISO2017}, without relying on any compiler extensions or 3rd-party
libraries, and is therefore compatible with any modern
standards-conforming C++17 compiler on any platform. Third, \emph{ease
  of use}: the package is extensively documented, and programmers of any
level should be able to use it right out of the box.

The fourth and final guiding principle is \emph{performance}: each and
every line of code in this library was carefully designed with maximum
performance in mind, and performance was tested and verified on a
variety of compilers and platforms. Indeed, the library was originally
designed for use in the author's own computationally-intensive
scientific computing projects, running both on high-end desktop/laptop
computers and high-performance computing nodes.

Other, more advanced multithreading libraries may offer more features
and/or higher performance. However, they typically consist of a vast
codebase with multiple components and dependencies, and involve complex
APIs that require a substantial time investment to learn. This library
is not intended to replace these more advanced libraries; instead, it
was designed for users who don't require very advanced features, and
prefer a simple and lightweight package that is easy to learn and use
and can be readily incorporated into existing or new projects.

\pagebreak

\hypertarget{overview-of-features}{%
\subsubsection{Overview of features}\label{overview-of-features}}

\begin{itemize}
\tightlist
\item
  \textbf{Fast:}

  \begin{itemize}
  \tightlist
  \item
    Built from scratch with maximum performance in mind.
  \item
    Suitable for use in high-performance computing nodes with a very
    large number of CPU cores.
  \item
    Compact code, to reduce both compilation time and binary size.
  \item
    Reusing threads avoids the overhead of creating and destroying them
    for individual tasks.
  \item
    A task queue ensures that there are never more threads running in
    parallel than allowed by the hardware.
  \end{itemize}
\item
  \textbf{Lightweight:}

  \begin{itemize}
  \tightlist
  \item
    Only \textsubscript{190} lines of code, excluding comments, blank
    lines, and the two optional helper classes.
  \item
    Single header file: simply
    \texttt{\#include\ "BS\_thread\_pool.hpp"} and you're all set!
  \item
    Header-only: no need to install or build the library.
  \item
    Self-contained: no external requirements or dependencies.
  \item
    Portable: uses only the C++ standard library, and works with any
    C++17-compliant compiler.
  \end{itemize}
\item
  \textbf{Easy to use:}

  \begin{itemize}
  \tightlist
  \item
    Very simple operation, using a handful of member functions.
  \item
    Every task submitted to the queue using the \texttt{submit()} member
    function automatically generates an \texttt{std::future}, which can
    be used to wait for the task to finish executing and/or obtain its
    eventual return value.
  \item
    Optionally, tasks may also be submitted using the
    \texttt{push\_task()} member function without generating a future,
    sacrificing convenience for even greater performance.
  \item
    The code is thoroughly documented using Doxygen comments - not only
    the interface, but also the implementation, in case the user would
    like to make modifications.
  \item
    The included test program \texttt{BS\_thread\_pool\_test.cpp} can be
    used to perform exhaustive automated tests and benchmarks, and also
    serves as a comprehensive example of how to properly use the
    package.
  \end{itemize}
\item
  \textbf{Helper classes:}

  \begin{itemize}
  \tightlist
  \item
    Automatically parallelize a loop into any number of parallel tasks
    using the \texttt{parallelize\_loop()} member function, and track
    its execution using the \texttt{BS::multi\_future} helper class.
  \item
    Synchronize output to a stream from multiple threads in parallel
    using the \texttt{BS::synced\_stream} helper class.
  \item
    Easily measure execution time for benchmarking purposes using the
    \texttt{BS::timer} helper class.
  \end{itemize}
\item
  \textbf{Additional features:}

  \begin{itemize}
  \tightlist
  \item
    Easily wait for all tasks in the queue to complete using the
    \texttt{wait\_for\_tasks()} member function.
  \item
    Change the number of threads in the pool safely and on-the-fly as
    needed using the \texttt{reset()} member function.
  \item
    Monitor the number of queued and/or running tasks using the
    \texttt{get\_tasks\_queued()}, \texttt{get\_tasks\_running()}, and
    \texttt{get\_tasks\_total()} member functions.
  \item
    Freely pause and resume the pool by modifying the \texttt{paused}
    member variable. When paused, threads do not retrieve new tasks out
    of the queue.
  \item
    Catch exceptions thrown by the submitted tasks.
  \item
    Under continuous and active development. Bug reports and feature
    requests are welcome, and should be made via
    \href{https://github.com/bshoshany/thread-pool/issues}{GitHub
    issues}.
  \end{itemize}
\end{itemize}

\hypertarget{compiling-and-compatibility}{%
\subsubsection{Compiling and
compatibility}\label{compiling-and-compatibility}}

This library should successfully compile on any C++17 standard-compliant
compiler, on all operating systems and architectures for which such a
compiler is available. Compatibility was verified with a 12-core /
24-thread AMD Ryzen 9 3900X CPU using the following compilers and
platforms:

\begin{itemize}
\tightlist
\item
  Windows 11 build 22000.675:

  \begin{itemize}
  \tightlist
  \item
    \href{https://gcc.gnu.org/}{GCC} v12.1.0
    (\href{https://winlibs.com/}{WinLibs build})
  \item
    \href{https://clang.llvm.org/}{Clang} v14.0.4
  \item
    \href{https://software.intel.com/content/www/us/en/develop/tools/oneapi/components/dpc-compiler.html}{Intel
    oneAPI C++ Compiler} v2022.1.0
  \item
    \href{https://docs.microsoft.com/en-us/cpp/}{MSVC} v19.32.31329
  \end{itemize}
\item
  Ubuntu 22.04 LTS:

  \begin{itemize}
  \tightlist
  \item
    \href{https://gcc.gnu.org/}{GCC} v12.0.1
  \item
    \href{https://clang.llvm.org/}{Clang} v14.0.0
  \end{itemize}
\end{itemize}

In addition, this library was tested on a
\href{https://www.computecanada.ca/}{Compute Canada} node equipped with
two 20-core / 40-thread Intel Xeon Gold 6148 CPUs (for a total of 40
cores and 80 threads), running CentOS Linux 7.9.2009, using
\href{https://gcc.gnu.org/}{GCC} v12.1.1.

The test program \texttt{BS\_thread\_pool\_test.cpp} was compiled
without warnings (with the warning flags
\texttt{-Wall\ -Wextra\ -Wconversion\ -Wsign-conversion\ -Wpedantic\ -Weffc++\ -Wshadow}
in GCC/Clang and \texttt{/W4} in MSVC), executed, and successfully
completed all \protect\hyperlink{testing-the-package}{automated tests}
and benchmarks using all of the compilers and systems mentioned above.

As this library requires C++17 features, the code must be compiled with
C++17 support:

\begin{itemize}
\tightlist
\item
  For GCC or Clang, use the \texttt{-std=c++17} flag. On Linux, you will
  also need to use the \texttt{-pthread} flag to enable the POSIX
  threads library.
\item
  For Intel, use \texttt{-std=c++17} on Linux or \texttt{/Qstd:c++17} on
  Windows.
\item
  For MSVC, use \texttt{/std:c++17}, and preferably also
  \texttt{/permissive-} to ensure standards conformance.
\end{itemize}

For maximum performance, it is recommended to compile with all available
compiler optimizations:

\begin{itemize}
\tightlist
\item
  For GCC or Clang, use the \texttt{-O3} flag.
\item
  For Intel, use \texttt{-O3} on Linux or \texttt{/O3} on Windows.
\item
  For MSVC, use \texttt{/O2}.
\end{itemize}

As an example, to compile the test program
\texttt{BS\_thread\_pool\_test.cpp} with warnings and optimizations, it
is recommended to use the following commands:

\begin{itemize}
\tightlist
\item
  On Windows with MSVC:
  \texttt{cl\ BS\_thread\_pool\_test.cpp\ /std:c++17\ /permissive-\ /O2\ /W4\ /EHsc\ /Fe:BS\_thread\_pool\_test.exe}
\item
  On Linux with GCC:
  \texttt{g++\ BS\_thread\_pool\_test.cpp\ -std=c++17\ -O3\ -Wall\ -Wextra\ -Wconversion\ -Wsign-conversion\ -Wpedantic\ -Weffc++\ -Wshadow\ -pthread\ -o\ BS\_thread\_pool\_test}
\end{itemize}

\hypertarget{installing-using-vcpkg}{%
\subsubsection{Installing using vcpkg}\label{installing-using-vcpkg}}

If you are using the \href{https://github.com/microsoft/vcpkg}{vcpkg}
C/C++ library manager, you can easily download and install this package
with the following commands.

On Linux/macOS:

\begin{Shaded}
\begin{Highlighting}[]
\NormalTok{./vcpkg install bshoshany{-}thread{-}pool}
\end{Highlighting}
\end{Shaded}

On Windows:

\begin{Shaded}
\begin{Highlighting}[]
\NormalTok{.\textbackslash{}vcpkg install bshoshany{-}thread{-}pool:x86{-}windows bshoshany{-}thread{-}pool:x64{-}windows}
\end{Highlighting}
\end{Shaded}

The thread pool will then be available automatically in the build system
you integrated vcpkg with (e.g. MSBuild or CMake). Simply write
\texttt{\#include\ "BS\_thread\_pool.hpp"} in any project to use the
thread pool, without having to copy to file into the project first. I
will update the vcpkg port with each new release, so it will be updated
automatically when you run \texttt{vcpkg\ upgrade}.

Please see the \href{https://github.com/microsoft/vcpkg}{vcpkg
repository} for more information on how to use vcpkg.

\hypertarget{getting-started}{%
\subsection{Getting started}\label{getting-started}}

\hypertarget{including-the-library}{%
\subsubsection{Including the library}\label{including-the-library}}

If you are not using a C++ library manager (such as vcpkg), simply
download the
\href{https://github.com/bshoshany/thread-pool/releases}{latest release}
from the GitHub repository, place the single header file
\texttt{BS\_thread\_pool.hpp} in the desired folder, and include it in
your program:

\begin{Shaded}
\begin{Highlighting}[]
\PreprocessorTok{\#include }\ImportTok{"BS\_thread\_pool.hpp"}
\end{Highlighting}
\end{Shaded}

The thread pool will now be accessible via the \texttt{BS::thread\_pool}
class.

\hypertarget{constructors}{%
\subsubsection{Constructors}\label{constructors}}

The default constructor creates a thread pool with as many threads as
the hardware can handle concurrently, as reported by the implementation
via \texttt{std::thread::hardware\_concurrency()}. This is usually
determined by the number of cores in the CPU. If a core is
hyperthreaded, it will count as two threads. For example:

\begin{Shaded}
\begin{Highlighting}[]
\CommentTok{// Constructs a thread pool with as many threads as available in the hardware.}
\NormalTok{BS}\OperatorTok{::}\NormalTok{thread\_pool pool}\OperatorTok{;}
\end{Highlighting}
\end{Shaded}

Optionally, a number of threads different from the hardware concurrency
can be specified as an argument to the constructor. However, note that
adding more threads than the hardware can handle will \textbf{not}
improve performance, and in fact will most likely hinder it. This option
exists in order to allow using \textbf{less} threads than the hardware
concurrency, in cases where you wish to leave some threads available for
other processes. For example:

\begin{Shaded}
\begin{Highlighting}[]
\CommentTok{// Constructs a thread pool with only 12 threads.}
\NormalTok{BS}\OperatorTok{::}\NormalTok{thread\_pool pool}\OperatorTok{(}\DecValTok{12}\OperatorTok{);}
\end{Highlighting}
\end{Shaded}

If your program's main thread only submits tasks to the thread pool and
waits for them to finish, and does not perform any computationally
intensive tasks on its own, then it is recommended to use the default
value for the number of threads. This ensures that all of the threads
available in the hardware will be put to work while the main thread
waits.

However, if your main thread does perform computationally intensive
tasks on its own, then it is recommended to use the value
\texttt{std::thread::hardware\_concurrency()\ -\ 1} for the number of
threads. In this case, the main thread plus the thread pool will
together take up exactly all the threads available in the hardware.

\hypertarget{getting-and-resetting-the-number-of-threads-in-the-pool}{%
\subsubsection{Getting and resetting the number of threads in the
pool}\label{getting-and-resetting-the-number-of-threads-in-the-pool}}

The member function \texttt{get\_thread\_count()} returns the number of
threads in the pool. This will be equal to
\texttt{std::thread::hardware\_concurrency()} if the default constructor
was used.

It is generally unnecessary to change the number of threads in the pool
after it has been created, since the whole point of a thread pool is
that you only create the threads once. However, if needed, this can be
done, safely and on-the-fly, using the \texttt{reset()} member function.

\texttt{reset()} will wait for all currently running tasks to be
completed, but will leave the rest of the tasks in the queue. Then it
will destroy the thread pool and create a new one with the desired new
number of threads, as specified in the function's argument (or the
hardware concurrency if no argument is given). The new thread pool will
then resume executing the tasks that remained in the queue and any new
submitted tasks.

\hypertarget{finding-the-version-of-the-package}{%
\subsubsection{Finding the version of the
package}\label{finding-the-version-of-the-package}}

If desired, the version of this package may be read during compilation
time from the macro \texttt{BS\_THREAD\_POOL\_VERSION}. The value will
be a string containing the version number and release date. For example:

\begin{Shaded}
\begin{Highlighting}[]
\BuiltInTok{std::}\NormalTok{cout}\OperatorTok{ \textless{}\textless{}} \StringTok{"Thread pool library version is "} \OperatorTok{\textless{}\textless{}}\NormalTok{ BS\_THREAD\_POOL\_VERSION }\OperatorTok{\textless{}\textless{}} \StringTok{".}\SpecialCharTok{\textbackslash{}n}\StringTok{"}\OperatorTok{;}
\end{Highlighting}
\end{Shaded}

Sample output:

\begin{Shaded}
\begin{Highlighting}[]
\NormalTok{Thread pool library version is v3.0.0 (2022{-}05{-}30).}
\end{Highlighting}
\end{Shaded}

This can be used, for example, to allow the same code to work with
several incompatible versions of the library.

\hypertarget{submitting-and-waiting-for-tasks}{%
\subsection{Submitting and waiting for
tasks}\label{submitting-and-waiting-for-tasks}}

\hypertarget{submitting-tasks-to-the-queue-with-futures}{%
\subsubsection{Submitting tasks to the queue with
futures}\label{submitting-tasks-to-the-queue-with-futures}}

A task can be any function, with zero or more arguments, and with or
without a return value. Once a task has been submitted to the queue, it
will be executed as soon as a thread becomes available. Tasks are
executed in the order that they were submitted (first-in, first-out).

The member function \texttt{submit()} is used to submit tasks to the
queue. The first argument is the function to execute, and the rest of
the arguments are the arguments to pass to the function, if any. The
return value is an \texttt{std::future} associated to the task. For
example:

\begin{Shaded}
\begin{Highlighting}[]
\CommentTok{// Submit a task without arguments to the queue, and get a future for it.}
\KeywordTok{auto}\NormalTok{ my\_future }\OperatorTok{=}\NormalTok{ pool}\OperatorTok{.}\NormalTok{submit}\OperatorTok{(}\NormalTok{task}\OperatorTok{);}
\CommentTok{// Submit a task with one argument to the queue, and get a future for it.}
\KeywordTok{auto}\NormalTok{ my\_future }\OperatorTok{=}\NormalTok{ pool}\OperatorTok{.}\NormalTok{submit}\OperatorTok{(}\NormalTok{task}\OperatorTok{,}\NormalTok{ arg}\OperatorTok{);}
\CommentTok{// Submit a task with two arguments to the queue, and get a future for it.}
\KeywordTok{auto}\NormalTok{ my\_future }\OperatorTok{=}\NormalTok{ pool}\OperatorTok{.}\NormalTok{submit}\OperatorTok{(}\NormalTok{task}\OperatorTok{,}\NormalTok{ arg1}\OperatorTok{,}\NormalTok{ arg2}\OperatorTok{);}
\end{Highlighting}
\end{Shaded}

If the submitted function has a return value of type \texttt{T}, then
the future will be of type
\texttt{std::future\textless{}T\textgreater{}}, and will be set to the
return value when the function finishes its execution. If the submitted
function does not have a return value, then the future will be an
\texttt{std::future\textless{}void\textgreater{}}, which will not return
any value but may still be used to wait for the function to finish.

Using \texttt{auto} for the return value of \texttt{submit()} means the
compiler will automatically detect which instance of the template
\texttt{std::future} to use. However, specifying the particular type
\texttt{std::future\textless{}T\textgreater{}}, as in the examples
below, is recommended for increased readability.

To wait until the task finishes, use the member function \texttt{wait()}
of the future. To obtain the return value, use the member function
\texttt{get()}, which will also automatically wait for the task to
finish if it hasn't yet. For example:

\begin{Shaded}
\begin{Highlighting}[]
\CommentTok{// Submit a task and get a future.}
\KeywordTok{auto}\NormalTok{ my\_future }\OperatorTok{=}\NormalTok{ pool}\OperatorTok{.}\NormalTok{submit}\OperatorTok{(}\NormalTok{task}\OperatorTok{);}
\CommentTok{// Do some other stuff while the task is executing.}
\NormalTok{do\_stuff}\OperatorTok{();}
\CommentTok{// Get the return value from the future, waiting for it to finish running if needed.}
\KeywordTok{auto}\NormalTok{ my\_return\_value }\OperatorTok{=}\NormalTok{ my\_future}\OperatorTok{.}\NormalTok{get}\OperatorTok{();}
\end{Highlighting}
\end{Shaded}

Here are some more concrete examples. The following program will print
out \texttt{42}:

\begin{Shaded}
\begin{Highlighting}[]
\PreprocessorTok{\#include }\ImportTok{"BS\_thread\_pool.hpp"}

\DataTypeTok{int}\NormalTok{ main}\OperatorTok{()}
\OperatorTok{\{}
\NormalTok{    BS}\OperatorTok{::}\NormalTok{thread\_pool pool}\OperatorTok{;}
    \BuiltInTok{std::}\NormalTok{future}\OperatorTok{\textless{}}\DataTypeTok{int}\OperatorTok{\textgreater{}}\NormalTok{ my\_future }\OperatorTok{=}\NormalTok{ pool}\OperatorTok{.}\NormalTok{submit}\OperatorTok{([]} \OperatorTok{\{} \ControlFlowTok{return} \DecValTok{42}\OperatorTok{;} \OperatorTok{\});}
    \BuiltInTok{std::}\NormalTok{cout}\OperatorTok{ \textless{}\textless{}}\NormalTok{ my\_future}\OperatorTok{.}\NormalTok{get}\OperatorTok{();}
\OperatorTok{\}}
\end{Highlighting}
\end{Shaded}

Here we used a
\href{https://en.cppreference.com/w/cpp/language/lambda}{lambda
expression} to quickly define the function on-the-fly. However, we can
also use a previously-defined function:

\begin{Shaded}
\begin{Highlighting}[]
\PreprocessorTok{\#include }\ImportTok{"BS\_thread\_pool.hpp"}

\DataTypeTok{int}\NormalTok{ the\_answer}\OperatorTok{()}
\OperatorTok{\{}
    \ControlFlowTok{return} \DecValTok{42}\OperatorTok{;}
\OperatorTok{\}}

\DataTypeTok{int}\NormalTok{ main}\OperatorTok{()}
\OperatorTok{\{}
\NormalTok{    BS}\OperatorTok{::}\NormalTok{thread\_pool pool}\OperatorTok{;}
    \BuiltInTok{std::}\NormalTok{future}\OperatorTok{\textless{}}\DataTypeTok{int}\OperatorTok{\textgreater{}}\NormalTok{ my\_future }\OperatorTok{=}\NormalTok{ pool}\OperatorTok{.}\NormalTok{submit}\OperatorTok{(}\NormalTok{the\_answer}\OperatorTok{);}
    \BuiltInTok{std::}\NormalTok{cout}\OperatorTok{ \textless{}\textless{}}\NormalTok{ my\_future}\OperatorTok{.}\NormalTok{get}\OperatorTok{();}
\OperatorTok{\}}
\end{Highlighting}
\end{Shaded}

The following is an example of submitting a function with arguments:

\begin{Shaded}
\begin{Highlighting}[]
\PreprocessorTok{\#include }\ImportTok{"BS\_thread\_pool.hpp"}

\DataTypeTok{int}\NormalTok{ multiply}\OperatorTok{(}\AttributeTok{const} \DataTypeTok{int}\NormalTok{ a}\OperatorTok{,} \AttributeTok{const} \DataTypeTok{int}\NormalTok{ b}\OperatorTok{)}
\OperatorTok{\{}
    \ControlFlowTok{return}\NormalTok{ a }\OperatorTok{*}\NormalTok{ b}\OperatorTok{;}
\OperatorTok{\}}

\DataTypeTok{int}\NormalTok{ main}\OperatorTok{()}
\OperatorTok{\{}
\NormalTok{    BS}\OperatorTok{::}\NormalTok{thread\_pool pool}\OperatorTok{;}
    \BuiltInTok{std::}\NormalTok{future}\OperatorTok{\textless{}}\DataTypeTok{int}\OperatorTok{\textgreater{}}\NormalTok{ my\_future }\OperatorTok{=}\NormalTok{ pool}\OperatorTok{.}\NormalTok{submit}\OperatorTok{(}\NormalTok{multiply}\OperatorTok{,} \DecValTok{6}\OperatorTok{,} \DecValTok{7}\OperatorTok{);}
    \BuiltInTok{std::}\NormalTok{cout}\OperatorTok{ \textless{}\textless{}}\NormalTok{ my\_future}\OperatorTok{.}\NormalTok{get}\OperatorTok{();}
\OperatorTok{\}}
\end{Highlighting}
\end{Shaded}

Finally, here is an example of submitting a function with no return
value and then using the future to wait for it to finish executing:

\begin{Shaded}
\begin{Highlighting}[]
\PreprocessorTok{\#include }\ImportTok{"BS\_thread\_pool.hpp"}

\DataTypeTok{void}\NormalTok{ sleep}\OperatorTok{()}
\OperatorTok{\{}
    \BuiltInTok{std::}\NormalTok{this\_thread}\BuiltInTok{::}\NormalTok{sleep\_for}\OperatorTok{(}\BuiltInTok{std::}\NormalTok{chrono}\BuiltInTok{::}\NormalTok{milliseconds}\OperatorTok{(}\DecValTok{1000}\OperatorTok{));}
\OperatorTok{\}}

\DataTypeTok{int}\NormalTok{ main}\OperatorTok{()}
\OperatorTok{\{}
\NormalTok{    BS}\OperatorTok{::}\NormalTok{thread\_pool pool}\OperatorTok{;}
    \BuiltInTok{std::}\NormalTok{future}\OperatorTok{\textless{}}\DataTypeTok{void}\OperatorTok{\textgreater{}}\NormalTok{ my\_future }\OperatorTok{=}\NormalTok{ pool}\OperatorTok{.}\NormalTok{submit}\OperatorTok{(}\NormalTok{sleep}\OperatorTok{);}
    \BuiltInTok{std::}\NormalTok{cout}\OperatorTok{ \textless{}\textless{}} \StringTok{"Waiting... "}\OperatorTok{;}
\NormalTok{    my\_future}\OperatorTok{.}\NormalTok{wait}\OperatorTok{();}
    \BuiltInTok{std::}\NormalTok{cout}\OperatorTok{ \textless{}\textless{}} \StringTok{"Done."}\OperatorTok{;}
\OperatorTok{\}}
\end{Highlighting}
\end{Shaded}

Here, the command
\texttt{std::this\_thread::sleep\_for(std::chrono::milliseconds(1000))}
instructs the thread to sleep for 1 second.

\hypertarget{submitting-tasks-to-the-queue-without-futures}{%
\subsubsection{Submitting tasks to the queue without
futures}\label{submitting-tasks-to-the-queue-without-futures}}

Usually, it is best to submit a task to the queue using
\texttt{submit()}. This allows you to wait for the task to finish and/or
get its return value later. However, sometimes a future is not needed,
for example when you just want to "set and forget" a certain task, or if
the task already communicates with the main thread or with other tasks
without using futures, such as via condition variables. In such cases,
you may wish to avoid the overhead involved in assigning a future to the
task in order to increase performance.

The member function \texttt{push\_task()} allows you to submit a task to
the queue without generating a future for it. The task can have any
number of arguments, but it cannot have a return value. For example:

\begin{Shaded}
\begin{Highlighting}[]
\CommentTok{// Submit a task without arguments or return value to the queue.}
\NormalTok{pool}\OperatorTok{.}\NormalTok{push\_task}\OperatorTok{(}\NormalTok{task}\OperatorTok{);}
\CommentTok{// Submit a task with one argument and no return value to the queue.}
\NormalTok{pool}\OperatorTok{.}\NormalTok{push\_task}\OperatorTok{(}\NormalTok{task}\OperatorTok{,}\NormalTok{ arg}\OperatorTok{);}
\CommentTok{// Submit a task with two arguments and no return value to the queue.}
\NormalTok{pool}\OperatorTok{.}\NormalTok{push\_task}\OperatorTok{(}\NormalTok{task}\OperatorTok{,}\NormalTok{ arg1}\OperatorTok{,}\NormalTok{ arg2}\OperatorTok{);}
\end{Highlighting}
\end{Shaded}

\hypertarget{manually-waiting-for-all-tasks-to-complete}{%
\subsubsection{Manually waiting for all tasks to
complete}\label{manually-waiting-for-all-tasks-to-complete}}

To wait for a \textbf{single} submitted task to complete, use
\texttt{submit()} and then use the \texttt{wait()} or \texttt{get()}
member functions of the obtained future. However, in cases where you
need to wait until \textbf{all} submitted tasks finish their execution,
or if the tasks have been submitted without futures using
\texttt{push\_task()}, you can use the member function
\texttt{wait\_for\_tasks()}.

Consider, for example, the following code:

\begin{Shaded}
\begin{Highlighting}[]
\PreprocessorTok{\#include }\ImportTok{"BS\_thread\_pool.hpp"}

\DataTypeTok{int}\NormalTok{ main}\OperatorTok{()}
\OperatorTok{\{}
\NormalTok{    BS}\OperatorTok{::}\NormalTok{thread\_pool pool}\OperatorTok{(}\DecValTok{5}\OperatorTok{);}
    \DataTypeTok{int}\NormalTok{ squares}\OperatorTok{[}\DecValTok{100}\OperatorTok{];}
    \ControlFlowTok{for} \OperatorTok{(}\DataTypeTok{int}\NormalTok{ i }\OperatorTok{=} \DecValTok{0}\OperatorTok{;}\NormalTok{ i }\OperatorTok{\textless{}} \DecValTok{100}\OperatorTok{;} \OperatorTok{++}\NormalTok{i}\OperatorTok{)}
\NormalTok{        pool}\OperatorTok{.}\NormalTok{push\_task}\OperatorTok{(}
            \OperatorTok{[\&}\NormalTok{squares}\OperatorTok{,}\NormalTok{ i}\OperatorTok{]}
            \OperatorTok{\{}
                \BuiltInTok{std::}\NormalTok{this\_thread}\BuiltInTok{::}\NormalTok{sleep\_for}\OperatorTok{(}\BuiltInTok{std::}\NormalTok{chrono}\BuiltInTok{::}\NormalTok{milliseconds}\OperatorTok{(}\DecValTok{50}\OperatorTok{));}
\NormalTok{                squares}\OperatorTok{[}\NormalTok{i}\OperatorTok{]} \OperatorTok{=}\NormalTok{ i }\OperatorTok{*}\NormalTok{ i}\OperatorTok{;}
            \OperatorTok{\});}
    \BuiltInTok{std::}\NormalTok{cout}\OperatorTok{ \textless{}\textless{}}\NormalTok{ squares}\OperatorTok{[}\DecValTok{50}\OperatorTok{];}
\OperatorTok{\}}
\end{Highlighting}
\end{Shaded}

The output will most likely be garbage, since the task that modifies
\texttt{squares{[}50{]}} has not yet finished executing by the time we
try to access that element - it's still waiting in the queue. One
solution would be to use \texttt{submit()} instead of
\texttt{push\_task()}, but perhaps we don't want the overhead of
generating 100 different futures. Instead, simply adding the line

\begin{Shaded}
\begin{Highlighting}[]
\NormalTok{pool}\OperatorTok{.}\NormalTok{wait\_for\_tasks}\OperatorTok{();}
\end{Highlighting}
\end{Shaded}

after the \texttt{for} loop will ensure - as efficiently as possible -
that all tasks have finished running before we attempt to access any
elements of the array \texttt{squares}, and the code will print out the
value \texttt{2500} as expected.

Note, however, that \texttt{wait\_for\_tasks()} will wait for
\textbf{all} the tasks in the queue, including those that are unrelated
to the \texttt{for} loop. Using
\protect\hyperlink{parallelizing-loops}{\texttt{parallelize\_loop()}}
would make much more sense in this particular case, as it will allow
waiting only for the tasks related to the loop.

\hypertarget{parallelizing-loops}{%
\subsubsection{Parallelizing loops}\label{parallelizing-loops}}

Consider the following loop:

\begin{Shaded}
\begin{Highlighting}[]
\ControlFlowTok{for} \OperatorTok{(}\NormalTok{T i }\OperatorTok{=}\NormalTok{ start}\OperatorTok{;}\NormalTok{ i }\OperatorTok{\textless{}}\NormalTok{ end}\OperatorTok{;} \OperatorTok{++}\NormalTok{i}\OperatorTok{)}
\NormalTok{    do\_something}\OperatorTok{(}\NormalTok{i}\OperatorTok{);}
\end{Highlighting}
\end{Shaded}

where:

\begin{itemize}
\tightlist
\item
  \texttt{T} is any signed or unsigned integer type.
\item
  The loop is over the range \texttt{{[}start,\ end)}, i.e. inclusive of
  \texttt{start} but exclusive of \texttt{end}.
\item
  \texttt{do\_something()} is an operation performed for each loop index
  \texttt{i}, such as modifying an array with \texttt{end\ -\ start}
  elements.
\end{itemize}

This loop may be automatically parallelized and submitted to the thread
pool's queue using the member function \texttt{parallelize\_loop()} as
follows:

\begin{Shaded}
\begin{Highlighting}[]
\KeywordTok{auto}\NormalTok{ loop }\OperatorTok{=} \OperatorTok{[](}\AttributeTok{const}\NormalTok{ T a}\OperatorTok{,} \AttributeTok{const}\NormalTok{ T b}\OperatorTok{)}
\OperatorTok{\{}
    \ControlFlowTok{for} \OperatorTok{(}\NormalTok{T i }\OperatorTok{=}\NormalTok{ a}\OperatorTok{;}\NormalTok{ i }\OperatorTok{\textless{}}\NormalTok{ b}\OperatorTok{;} \OperatorTok{++}\NormalTok{i}\OperatorTok{)}
\NormalTok{        do\_something}\OperatorTok{(}\NormalTok{i}\OperatorTok{);}
\OperatorTok{\};}
\NormalTok{BS}\OperatorTok{::}\NormalTok{multi\_future}\OperatorTok{\textless{}}\DataTypeTok{void}\OperatorTok{\textgreater{}}\NormalTok{ loop\_future }\OperatorTok{=}\NormalTok{ pool}\OperatorTok{.}\NormalTok{parallelize\_loop}\OperatorTok{(}\NormalTok{start}\OperatorTok{,}\NormalTok{ end}\OperatorTok{,}\NormalTok{ loop}\OperatorTok{,}\NormalTok{ n}\OperatorTok{);}
\NormalTok{loop\_future}\OperatorTok{.}\NormalTok{wait}\OperatorTok{();}
\end{Highlighting}
\end{Shaded}

Here's how this works:

\begin{itemize}
\tightlist
\item
  The lambda function \texttt{loop()} takes two indices, \texttt{a}, and
  \texttt{b}, and executes only the portion of the loop in the range
  \texttt{{[}a,\ b)}.

  \begin{itemize}
  \tightlist
  \item
    Note that this lambda was defined here separately for clarity. In
    practice, the lambda function will usually be defined within the
    call to \texttt{parallelize\_loop()} itself, as in the examples
    below.
  \item
    \texttt{loop()} can also be an ordinary function (with or without a
    return value) instead of a lambda function, but that may be less
    useful, since typically one would like to capture some of the
    surrounding variables.
  \end{itemize}
\item
  When \texttt{parallelize\_loop(start,\ end,\ loop,\ n)} is called, it
  will divide the range of indices \texttt{{[}start,\ end)} into
  \texttt{n} blocks of the form \texttt{{[}a,\ b)}. For example, if the
  range is \texttt{{[}0,\ 9)} and there are 3 blocks, then the blocks
  will be the ranges \texttt{{[}0,\ 3)}, \texttt{{[}3,\ 6)}, and
  \texttt{{[}6,\ 9)}. If possible, the blocks will be equal in size,
  otherwise the last block may be a bit longer.
\item
  Then, a task will be submitted for each block, consisting of the
  function \texttt{loop()} with its two arguments being the start and
  end of the range \texttt{{[}a,\ b)} of each block.
\item
  Each task will have an
  \texttt{std::future\textless{}void\textgreater{}} assigned to it, and
  all these futures will be stored inside an object
  \texttt{loop\_future} of the helper class template
  \texttt{BS::multi\_future}.
\item
  When \texttt{loop\_future.wait()} is called, the main thread will wait
  until all tasks generated by \texttt{parallelize\_loop()} finish
  executing, and only those tasks - not any other tasks that also happen
  to be in the queue. This is essentially the role of the
  \texttt{BS::multi\_future} class: to wait for a specific group of
  tasks, in this case the tasks running the loop blocks.
\end{itemize}

If the fourth argument \texttt{n} is not specified, the number of blocks
will be equal to the number of threads in the pool. For best
performance, it is recommended to do your own benchmarks to find the
optimal number of blocks for each loop (you can use the
\texttt{BS::timer} helper class - see
\protect\hyperlink{measuring-execution-time}{below}). Using less tasks
than there are threads may be preferred if you are also running other
tasks in parallel. Using more tasks than there are threads may improve
performance in some cases.

As a simple example, the following code will calculate the squares of
all integers from 0 to 99. Since there are 10 threads, and we did not
specify a fourth argument, the loop will be divided into 10 blocks, each
calculating 10 squares:

\begin{Shaded}
\begin{Highlighting}[]
\PreprocessorTok{\#include }\ImportTok{"BS\_thread\_pool.hpp"}

\DataTypeTok{int}\NormalTok{ main}\OperatorTok{()}
\OperatorTok{\{}
\NormalTok{    BS}\OperatorTok{::}\NormalTok{thread\_pool pool}\OperatorTok{(}\DecValTok{10}\OperatorTok{);}
    \DataTypeTok{int}\NormalTok{ squares}\OperatorTok{[}\DecValTok{100}\OperatorTok{];}
\NormalTok{    pool}\OperatorTok{.}\NormalTok{parallelize\_loop}\OperatorTok{(}\DecValTok{0}\OperatorTok{,} \DecValTok{100}\OperatorTok{,}
                          \OperatorTok{[\&}\NormalTok{squares}\OperatorTok{](}\AttributeTok{const} \DataTypeTok{int}\NormalTok{ a}\OperatorTok{,} \AttributeTok{const} \DataTypeTok{int}\NormalTok{ b}\OperatorTok{)}
                          \OperatorTok{\{}
                              \ControlFlowTok{for} \OperatorTok{(}\DataTypeTok{int}\NormalTok{ i }\OperatorTok{=}\NormalTok{ a}\OperatorTok{;}\NormalTok{ i }\OperatorTok{\textless{}}\NormalTok{ b}\OperatorTok{;} \OperatorTok{++}\NormalTok{i}\OperatorTok{)}
\NormalTok{                                  squares}\OperatorTok{[}\NormalTok{i}\OperatorTok{]} \OperatorTok{=}\NormalTok{ i }\OperatorTok{*}\NormalTok{ i}\OperatorTok{;}
                          \OperatorTok{\})}
        \OperatorTok{.}\NormalTok{wait}\OperatorTok{();}
    \BuiltInTok{std::}\NormalTok{cout}\OperatorTok{ \textless{}\textless{}}\NormalTok{ squares}\OperatorTok{[}\DecValTok{50}\OperatorTok{];}
\OperatorTok{\}}
\end{Highlighting}
\end{Shaded}

Note that here, for simplicity, instead of creating a
\texttt{BS::multi\_future} and then using it to wait, we simply called
the \texttt{wait()} member function directly on the temporary object
returned by \texttt{parallelize\_loop()}. This is a convenient shortcut
when we have nothing else to do while waiting.

\hypertarget{loops-with-return-values}{%
\subsubsection{Loops with return
values}\label{loops-with-return-values}}

Usually, \texttt{parallelize\_loop()} should take functions with no
return values. This is because the function will be executed once for
each block, but the blocks are managed by the thread pool, so there's
limited usability in returning one value per block. However, for the
case where this is desired, such as for summation or some sorting
algorithms, \texttt{parallelize\_loop()} does accept functions with
return values, in which case it returns a
\texttt{BS::multi\_future\textless{}T\textgreater{}} object where
\texttt{T} is the return value.

Here's an example of summing all the numbers from 1 to 100:

\begin{Shaded}
\begin{Highlighting}[]
\PreprocessorTok{\#include }\ImportTok{"BS\_thread\_pool.hpp"}

\DataTypeTok{int}\NormalTok{ main}\OperatorTok{()}
\OperatorTok{\{}
\NormalTok{    BS}\OperatorTok{::}\NormalTok{thread\_pool pool}\OperatorTok{;}
\NormalTok{    BS}\OperatorTok{::}\NormalTok{multi\_future}\OperatorTok{\textless{}}\DataTypeTok{int}\OperatorTok{\textgreater{}}\NormalTok{ mf }\OperatorTok{=}\NormalTok{ pool}\OperatorTok{.}\NormalTok{parallelize\_loop}\OperatorTok{(}\DecValTok{1}\OperatorTok{,} \DecValTok{101}\OperatorTok{,}
                                                     \OperatorTok{[](}\AttributeTok{const} \DataTypeTok{int}\NormalTok{ a}\OperatorTok{,} \AttributeTok{const} \DataTypeTok{int}\NormalTok{ b}\OperatorTok{)}
                                                     \OperatorTok{\{}
                                                         \DataTypeTok{int}\NormalTok{ block\_total }\OperatorTok{=} \DecValTok{0}\OperatorTok{;}
                                                         \ControlFlowTok{for} \OperatorTok{(}\DataTypeTok{int}\NormalTok{ i }\OperatorTok{=}\NormalTok{ a}\OperatorTok{;}\NormalTok{ i }\OperatorTok{\textless{}}\NormalTok{ b}\OperatorTok{;} \OperatorTok{++}\NormalTok{i}\OperatorTok{)}
\NormalTok{                                                             block\_total }\OperatorTok{+=}\NormalTok{ i}\OperatorTok{;}
                                                         \ControlFlowTok{return}\NormalTok{ block\_total}\OperatorTok{;}
                                                     \OperatorTok{\});}
    \BuiltInTok{std::}\NormalTok{vector}\OperatorTok{\textless{}}\DataTypeTok{int}\OperatorTok{\textgreater{}}\NormalTok{ totals }\OperatorTok{=}\NormalTok{ mf}\OperatorTok{.}\NormalTok{get}\OperatorTok{();}
    \DataTypeTok{int}\NormalTok{ sum }\OperatorTok{=} \DecValTok{0}\OperatorTok{;}
    \ControlFlowTok{for} \OperatorTok{(}\AttributeTok{const} \DataTypeTok{int}\NormalTok{ t }\OperatorTok{:}\NormalTok{ totals}\OperatorTok{)}
\NormalTok{        sum }\OperatorTok{+=}\NormalTok{ t}\OperatorTok{;}
    \BuiltInTok{std::}\NormalTok{cout}\OperatorTok{ \textless{}\textless{}}\NormalTok{ sum}\OperatorTok{;}
\OperatorTok{\}}
\end{Highlighting}
\end{Shaded}

Note that calling \texttt{get()} on a
\texttt{BS::multi\_future\textless{}T\textgreater{}} object returns an
\texttt{std::vector\textless{}T\textgreater{}} with the values obtained
from each future. In this case, the values will be the partial sums from
each block, so when we add them up, we will get the total sum.

\hypertarget{helper-classes}{%
\subsection{Helper classes}\label{helper-classes}}

\hypertarget{handling-multiple-futures-at-once}{%
\subsubsection{Handling multiple futures at
once}\label{handling-multiple-futures-at-once}}

The helper class template
\texttt{BS::multi\_future\textless{}T\textgreater{}}, already introduced
in the context of \texttt{parallelize\_loop()}, provides a convenient
way to collect and access groups of futures. The futures are stored in a
public member variable \texttt{f} of type
\texttt{std::vector\textless{}std::future\textless{}T\textgreater{}\textgreater{}},
so all standard \texttt{std::vector} operations are available for
organizing the futures. Once the futures are stored, you can use
\texttt{wait()} to wait for all of them at once or \texttt{get()} to get
an \texttt{std::vector\textless{}T\textgreater{}} with the results from
all of them. Here's a simple example:

\begin{Shaded}
\begin{Highlighting}[]
\PreprocessorTok{\#include }\ImportTok{"BS\_thread\_pool.hpp"}

\DataTypeTok{int}\NormalTok{ square}\OperatorTok{(}\AttributeTok{const} \DataTypeTok{int}\NormalTok{ i}\OperatorTok{)}
\OperatorTok{\{}
    \BuiltInTok{std::}\NormalTok{this\_thread}\BuiltInTok{::}\NormalTok{sleep\_for}\OperatorTok{(}\BuiltInTok{std::}\NormalTok{chrono}\BuiltInTok{::}\NormalTok{milliseconds}\OperatorTok{(}\DecValTok{500}\OperatorTok{));}
    \ControlFlowTok{return}\NormalTok{ i }\OperatorTok{*}\NormalTok{ i}\OperatorTok{;}
\OperatorTok{\};}

\DataTypeTok{int}\NormalTok{ main}\OperatorTok{()}
\OperatorTok{\{}
\NormalTok{    BS}\OperatorTok{::}\NormalTok{thread\_pool pool}\OperatorTok{;}
\NormalTok{    BS}\OperatorTok{::}\NormalTok{multi\_future}\OperatorTok{\textless{}}\DataTypeTok{int}\OperatorTok{\textgreater{}}\NormalTok{ mf1}\OperatorTok{;}
\NormalTok{    BS}\OperatorTok{::}\NormalTok{multi\_future}\OperatorTok{\textless{}}\DataTypeTok{int}\OperatorTok{\textgreater{}}\NormalTok{ mf2}\OperatorTok{;}
    \ControlFlowTok{for} \OperatorTok{(}\DataTypeTok{int}\NormalTok{ i }\OperatorTok{=} \DecValTok{0}\OperatorTok{;}\NormalTok{ i }\OperatorTok{\textless{}} \DecValTok{100}\OperatorTok{;} \OperatorTok{++}\NormalTok{i}\OperatorTok{)}
\NormalTok{        mf1}\OperatorTok{.}\NormalTok{f}\OperatorTok{.}\NormalTok{push\_back}\OperatorTok{(}\NormalTok{pool}\OperatorTok{.}\NormalTok{submit}\OperatorTok{(}\NormalTok{square}\OperatorTok{,}\NormalTok{ i}\OperatorTok{));}
    \ControlFlowTok{for} \OperatorTok{(}\DataTypeTok{int}\NormalTok{ i }\OperatorTok{=} \DecValTok{100}\OperatorTok{;}\NormalTok{ i }\OperatorTok{\textless{}} \DecValTok{200}\OperatorTok{;} \OperatorTok{++}\NormalTok{i}\OperatorTok{)}
\NormalTok{        mf2}\OperatorTok{.}\NormalTok{f}\OperatorTok{.}\NormalTok{push\_back}\OperatorTok{(}\NormalTok{pool}\OperatorTok{.}\NormalTok{submit}\OperatorTok{(}\NormalTok{square}\OperatorTok{,}\NormalTok{ i}\OperatorTok{));}
    \CommentTok{/// ...}
    \CommentTok{/// Do some stuff while the first group of tasks executes...}
    \CommentTok{/// ...}
    \AttributeTok{const} \BuiltInTok{std::}\NormalTok{vector}\OperatorTok{\textless{}}\DataTypeTok{int}\OperatorTok{\textgreater{}}\NormalTok{ squares1 }\OperatorTok{=}\NormalTok{ mf1}\OperatorTok{.}\NormalTok{get}\OperatorTok{();}
    \BuiltInTok{std::}\NormalTok{cout}\OperatorTok{ \textless{}\textless{}} \StringTok{"Results from the first group:"} \OperatorTok{\textless{}\textless{}} \CharTok{\textquotesingle{}}\SpecialCharTok{\textbackslash{}n}\CharTok{\textquotesingle{}}\OperatorTok{;}
    \ControlFlowTok{for} \OperatorTok{(}\AttributeTok{const} \DataTypeTok{int}\NormalTok{ s }\OperatorTok{:}\NormalTok{ squares1}\OperatorTok{)}
        \BuiltInTok{std::}\NormalTok{cout}\OperatorTok{ \textless{}\textless{}}\NormalTok{ s }\OperatorTok{\textless{}\textless{}} \CharTok{\textquotesingle{} \textquotesingle{}}\OperatorTok{;}
    \CommentTok{/// ...}
    \CommentTok{/// Do other stuff while the second group of tasks executes...}
    \CommentTok{/// ...}
    \AttributeTok{const} \BuiltInTok{std::}\NormalTok{vector}\OperatorTok{\textless{}}\DataTypeTok{int}\OperatorTok{\textgreater{}}\NormalTok{ squares2 }\OperatorTok{=}\NormalTok{ mf2}\OperatorTok{.}\NormalTok{get}\OperatorTok{();}
    \BuiltInTok{std::}\NormalTok{cout}\OperatorTok{ \textless{}\textless{}} \CharTok{\textquotesingle{}}\SpecialCharTok{\textbackslash{}n}\CharTok{\textquotesingle{}} \OperatorTok{\textless{}\textless{}} \StringTok{"Results from the second group:"} \OperatorTok{\textless{}\textless{}} \CharTok{\textquotesingle{}}\SpecialCharTok{\textbackslash{}n}\CharTok{\textquotesingle{}}\OperatorTok{;}
    \ControlFlowTok{for} \OperatorTok{(}\AttributeTok{const} \DataTypeTok{int}\NormalTok{ s }\OperatorTok{:}\NormalTok{ squares2}\OperatorTok{)}
        \BuiltInTok{std::}\NormalTok{cout}\OperatorTok{ \textless{}\textless{}}\NormalTok{ s }\OperatorTok{\textless{}\textless{}} \CharTok{\textquotesingle{} \textquotesingle{}}\OperatorTok{;}
\OperatorTok{\}}
\end{Highlighting}
\end{Shaded}

In this example, we simulate complicated tasks by having each task wait
for 500ms before returning its result. We collect the futures of the
tasks submitted within each loop into two separate
\texttt{BS::multi\_future\textless{}int\textgreater{}} objects.
\texttt{mf1} holds the results from the first loop, and \texttt{mf2}
holds the results from the second loop. Now we can wait for and/or get
the results from \texttt{mf1} whenever is convenient, and separately
wait for and/or get the results from \texttt{mf2} at another time.

\hypertarget{synchronizing-printing-to-an-output-stream}{%
\subsubsection{Synchronizing printing to an output
stream}\label{synchronizing-printing-to-an-output-stream}}

When printing to an output stream from multiple threads in parallel, the
output may become garbled. For example, consider this code:

\begin{Shaded}
\begin{Highlighting}[]
\PreprocessorTok{\#include }\ImportTok{"BS\_thread\_pool.hpp"}

\DataTypeTok{int}\NormalTok{ main}\OperatorTok{()}
\OperatorTok{\{}
\NormalTok{    BS}\OperatorTok{::}\NormalTok{thread\_pool pool}\OperatorTok{;}
    \ControlFlowTok{for} \OperatorTok{(}\DataTypeTok{size\_t}\NormalTok{ i }\OperatorTok{=} \DecValTok{1}\OperatorTok{;}\NormalTok{ i }\OperatorTok{\textless{}=} \DecValTok{5}\OperatorTok{;} \OperatorTok{++}\NormalTok{i}\OperatorTok{)}
\NormalTok{        pool}\OperatorTok{.}\NormalTok{push\_task}\OperatorTok{([}\NormalTok{i}\OperatorTok{]} \OperatorTok{\{} \BuiltInTok{std::}\NormalTok{cout}\OperatorTok{ \textless{}\textless{}} \StringTok{"Task no. "} \OperatorTok{\textless{}\textless{}}\NormalTok{ i }\OperatorTok{\textless{}\textless{}} \StringTok{" executing.}\SpecialCharTok{\textbackslash{}n}\StringTok{"}\OperatorTok{;} \OperatorTok{\});}
\OperatorTok{\}}
\end{Highlighting}
\end{Shaded}

The output may look as follows:

\begin{Shaded}
\begin{Highlighting}[]
\NormalTok{Task no. Task no. 2Task no. 5 executing.}
\NormalTok{Task no.  executing.}
\NormalTok{Task no. 1 executing.}
\NormalTok{4 executing.}
\NormalTok{3 executing.}
\end{Highlighting}
\end{Shaded}

The reason is that, although each \textbf{individual} insertion to
\texttt{std::cout} is thread-safe, there is no mechanism in place to
ensure subsequent insertions from the same thread are printed
contiguously.

The helper class \texttt{BS::synced\_stream} is designed to eliminate
such synchronization issues. The constructor takes one optional
argument, specifying the output stream to print to. If no argument is
supplied, \texttt{std::cout} will be used:

\begin{Shaded}
\begin{Highlighting}[]
\CommentTok{// Construct a synced stream that will print to std::cout.}
\NormalTok{BS}\OperatorTok{::}\NormalTok{synced\_stream sync\_out}\OperatorTok{;}
\CommentTok{// Construct a synced stream that will print to the output stream my\_stream.}
\NormalTok{BS}\OperatorTok{::}\NormalTok{synced\_stream sync\_out}\OperatorTok{(}\NormalTok{my\_stream}\OperatorTok{);}
\end{Highlighting}
\end{Shaded}

The member function \texttt{print()} takes an arbitrary number of
arguments, which are inserted into the stream one by one, in the order
they were given. \texttt{println()} does the same, but also prints a
newline character \texttt{\textbackslash{}n} at the end, for
convenience. A mutex is used to synchronize this process, so that any
other calls to \texttt{print()} or \texttt{println()} using the same
\texttt{BS::synced\_stream} object must wait until the previous call has
finished.

As an example, this code:

\begin{Shaded}
\begin{Highlighting}[]
\PreprocessorTok{\#include }\ImportTok{"BS\_thread\_pool.hpp"}

\DataTypeTok{int}\NormalTok{ main}\OperatorTok{()}
\OperatorTok{\{}
\NormalTok{    BS}\OperatorTok{::}\NormalTok{synced\_stream sync\_out}\OperatorTok{;}
\NormalTok{    BS}\OperatorTok{::}\NormalTok{thread\_pool pool}\OperatorTok{;}
    \ControlFlowTok{for} \OperatorTok{(}\DataTypeTok{size\_t}\NormalTok{ i }\OperatorTok{=} \DecValTok{1}\OperatorTok{;}\NormalTok{ i }\OperatorTok{\textless{}=} \DecValTok{5}\OperatorTok{;} \OperatorTok{++}\NormalTok{i}\OperatorTok{)}
\NormalTok{        pool}\OperatorTok{.}\NormalTok{push\_task}\OperatorTok{([}\NormalTok{i}\OperatorTok{,} \OperatorTok{\&}\NormalTok{sync\_out}\OperatorTok{]}
                       \OperatorTok{\{}
                       \NormalTok{    sync\_out}\OperatorTok{.}\NormalTok{println}\OperatorTok{(}\StringTok{"Task no. "}\OperatorTok{,}\NormalTok{ i}\OperatorTok{,} \StringTok{" executing."}\OperatorTok{);}
                       \OperatorTok{\});}
\OperatorTok{\}}
\end{Highlighting}
\end{Shaded}

Will print out:

\begin{Shaded}
\begin{Highlighting}[]
\NormalTok{Task no. 1 executing.}
\NormalTok{Task no. 2 executing.}
\NormalTok{Task no. 3 executing.}
\NormalTok{Task no. 4 executing.}
\NormalTok{Task no. 5 executing.}
\end{Highlighting}
\end{Shaded}

\textbf{Warning:} Always create the \texttt{BS::synced\_stream} object
\textbf{before} the \texttt{BS::thread\_pool} object, as we did in this
example. When the \texttt{BS::thread\_pool} object goes out of scope, it
waits for the remaining tasks to be executed. If the
\texttt{BS::synced\_stream} object goes out of scope before the
\texttt{BS::thread\_pool} object, then any tasks using the
\texttt{BS::synced\_stream} will crash. Since objects are destructed in
the opposite order of construction, creating the
\texttt{BS::synced\_stream} object before the \texttt{BS::thread\_pool}
object ensures that the \texttt{BS::synced\_stream} is always available
to the tasks, even while the pool is destructing.

\hypertarget{measuring-execution-time}{%
\subsubsection{Measuring execution
time}\label{measuring-execution-time}}

If you are using a thread pool, then your code is most likely
performance-critical. Achieving maximum performance requires performing
a considerable amount of benchmarking to determine the optimal settings
and algorithms. Therefore, it is important to be able to measure the
execution time of various computations and operations under different
conditions.

The helper class \texttt{BS::timer} provides a simple way to measure
execution time. It is very straightforward to use:

\begin{enumerate}
\tightlist
\item
  Create a new \texttt{BS::timer} object.
\item
  Immediately before you execute the computation that you want to time,
  call the \texttt{start()} member function.
\item
  Immediately after the computation ends, call the \texttt{stop()}
  member function.
\item
  Use the member function \texttt{ms()} to obtain the elapsed time for
  the computation in milliseconds.
\end{enumerate}

For example:

\begin{Shaded}
\begin{Highlighting}[]
\NormalTok{BS}\OperatorTok{::}\NormalTok{timer tmr}\OperatorTok{;}
\NormalTok{tmr}\OperatorTok{.}\NormalTok{start}\OperatorTok{();}
\NormalTok{do\_something}\OperatorTok{();}
\NormalTok{tmr}\OperatorTok{.}\NormalTok{stop}\OperatorTok{();}
\BuiltInTok{std::}\NormalTok{cout}\OperatorTok{ \textless{}\textless{}} \StringTok{"The elapsed time was "} \OperatorTok{\textless{}\textless{}}\NormalTok{ tmr}\OperatorTok{.}\NormalTok{ms}\OperatorTok{()} \OperatorTok{\textless{}\textless{}} \StringTok{" ms.}\SpecialCharTok{\textbackslash{}n}\StringTok{"}\OperatorTok{;}
\end{Highlighting}
\end{Shaded}

A practical application of the \texttt{BS::timer} class can be found in
the benchmark portion of the test program
\texttt{BS\_thread\_pool\_test.cpp}.

\hypertarget{other-features}{%
\subsection{Other features}\label{other-features}}

\hypertarget{monitoring-the-tasks}{%
\subsubsection{Monitoring the tasks}\label{monitoring-the-tasks}}

Sometimes you may wish to monitor what is happening with the tasks you
submitted to the pool. This may be done using three member functions:

\begin{itemize}
\tightlist
\item
  \texttt{get\_tasks\_queued()} gets the number of tasks currently
  waiting in the queue to be executed by the threads.
\item
  \texttt{get\_tasks\_running()} gets the number of tasks currently
  being executed by the threads.
\item
  \texttt{get\_tasks\_total()} gets the total number of unfinished
  tasks: either still in the queue, or running in a thread.
\item
  Note that
  \texttt{get\_tasks\_total()\ ==\ get\_tasks\_queued()\ +\ get\_tasks\_running()}.
\end{itemize}

These functions are demonstrated in the following program:

\begin{Shaded}
\begin{Highlighting}[]
\PreprocessorTok{\#include }\ImportTok{"BS\_thread\_pool.hpp"}

\NormalTok{BS}\OperatorTok{::}\NormalTok{synced\_stream sync\_out}\OperatorTok{;}
\NormalTok{BS}\OperatorTok{::}\NormalTok{thread\_pool pool}\OperatorTok{(}\DecValTok{4}\OperatorTok{);}

\DataTypeTok{void}\NormalTok{ sleep\_half\_second}\OperatorTok{(}\AttributeTok{const} \DataTypeTok{size\_t}\NormalTok{ i}\OperatorTok{)}
\OperatorTok{\{}
    \BuiltInTok{std::}\NormalTok{this\_thread}\BuiltInTok{::}\NormalTok{sleep\_for}\OperatorTok{(}\BuiltInTok{std::}\NormalTok{chrono}\BuiltInTok{::}\NormalTok{milliseconds}\OperatorTok{(}\DecValTok{500}\OperatorTok{));}
\NormalTok{    sync\_out}\OperatorTok{.}\NormalTok{println}\OperatorTok{(}\StringTok{"Task "}\OperatorTok{,}\NormalTok{ i}\OperatorTok{,} \StringTok{" done."}\OperatorTok{);}
\OperatorTok{\}}

\DataTypeTok{void}\NormalTok{ monitor\_tasks}\OperatorTok{()}
\OperatorTok{\{}
\NormalTok{    sync\_out}\OperatorTok{.}\NormalTok{println}\OperatorTok{(}\NormalTok{pool}\OperatorTok{.}\NormalTok{get\_tasks\_total}\OperatorTok{(),} \StringTok{" tasks total, "}\OperatorTok{,}
                     \NormalTok{pool}\OperatorTok{.}\NormalTok{get\_tasks\_running}\OperatorTok{(),} \StringTok{" tasks running, "}\OperatorTok{,}
                     \NormalTok{pool}\OperatorTok{.}\NormalTok{get\_tasks\_queued}\OperatorTok{(),} \StringTok{" tasks queued."}\OperatorTok{);}
\OperatorTok{\}}

\DataTypeTok{int}\NormalTok{ main}\OperatorTok{()}
\OperatorTok{\{}
    \ControlFlowTok{for} \OperatorTok{(}\DataTypeTok{size\_t}\NormalTok{ i }\OperatorTok{=} \DecValTok{0}\OperatorTok{;}\NormalTok{ i }\OperatorTok{\textless{}} \DecValTok{12}\OperatorTok{;} \OperatorTok{++}\NormalTok{i}\OperatorTok{)}
\NormalTok{        pool}\OperatorTok{.}\NormalTok{push\_task}\OperatorTok{(}\NormalTok{sleep\_half\_second}\OperatorTok{,}\NormalTok{ i}\OperatorTok{);}
\NormalTok{    monitor\_tasks}\OperatorTok{();}
    \BuiltInTok{std::}\NormalTok{this\_thread}\BuiltInTok{::}\NormalTok{sleep\_for}\OperatorTok{(}\BuiltInTok{std::}\NormalTok{chrono}\BuiltInTok{::}\NormalTok{milliseconds}\OperatorTok{(}\DecValTok{750}\OperatorTok{));}
\NormalTok{    monitor\_tasks}\OperatorTok{();}
    \BuiltInTok{std::}\NormalTok{this\_thread}\BuiltInTok{::}\NormalTok{sleep\_for}\OperatorTok{(}\BuiltInTok{std::}\NormalTok{chrono}\BuiltInTok{::}\NormalTok{milliseconds}\OperatorTok{(}\DecValTok{500}\OperatorTok{));}
\NormalTok{    monitor\_tasks}\OperatorTok{();}
    \BuiltInTok{std::}\NormalTok{this\_thread}\BuiltInTok{::}\NormalTok{sleep\_for}\OperatorTok{(}\BuiltInTok{std::}\NormalTok{chrono}\BuiltInTok{::}\NormalTok{milliseconds}\OperatorTok{(}\DecValTok{500}\OperatorTok{));}
\NormalTok{    monitor\_tasks}\OperatorTok{();}
\OperatorTok{\}}
\end{Highlighting}
\end{Shaded}

Assuming you have at least 4 hardware threads (so that 4 tasks can run
concurrently), the output should be similar to:

\begin{Shaded}
\begin{Highlighting}[]
\NormalTok{12 tasks total, 0 tasks running, 12 tasks queued.}
\NormalTok{Task 0 done.}
\NormalTok{Task 1 done.}
\NormalTok{Task 2 done.}
\NormalTok{Task 3 done.}
\NormalTok{8 tasks total, 4 tasks running, 4 tasks queued.}
\NormalTok{Task 4 done.}
\NormalTok{Task 5 done.}
\NormalTok{Task 6 done.}
\NormalTok{Task 7 done.}
\NormalTok{4 tasks total, 4 tasks running, 0 tasks queued.}
\NormalTok{Task 8 done.}
\NormalTok{Task 9 done.}
\NormalTok{Task 10 done.}
\NormalTok{Task 11 done.}
\NormalTok{0 tasks total, 0 tasks running, 0 tasks queued.}
\end{Highlighting}
\end{Shaded}

\hypertarget{pausing-the-workers}{%
\subsubsection{Pausing the workers}\label{pausing-the-workers}}

Sometimes you may wish to temporarily pause the execution of tasks, or
perhaps you want to submit tasks to the queue in advance and only start
executing them at a later time. You can do this using the public member
variable \texttt{paused}.

When \texttt{paused} is set to \texttt{true}, the workers will
temporarily stop retrieving new tasks out of the queue. However, any
tasks already executed will keep running until they are done, since the
thread pool has no control over the internal code of your tasks. If you
need to pause a task in the middle of its execution, you must do that
manually by programming your own pause mechanism into the task itself.
To resume retrieving tasks, set \texttt{paused} back to its default
value of \texttt{false}.

Here is an example:

\begin{Shaded}
\begin{Highlighting}[]
\PreprocessorTok{\#include }\ImportTok{"BS\_thread\_pool.hpp"}

\NormalTok{BS}\OperatorTok{::}\NormalTok{synced\_stream sync\_out}\OperatorTok{;}
\NormalTok{BS}\OperatorTok{::}\NormalTok{thread\_pool pool}\OperatorTok{(}\DecValTok{4}\OperatorTok{);}

\DataTypeTok{void}\NormalTok{ sleep\_half\_second}\OperatorTok{(}\AttributeTok{const} \DataTypeTok{size\_t}\NormalTok{ i}\OperatorTok{)}
\OperatorTok{\{}
    \BuiltInTok{std::}\NormalTok{this\_thread}\BuiltInTok{::}\NormalTok{sleep\_for}\OperatorTok{(}\BuiltInTok{std::}\NormalTok{chrono}\BuiltInTok{::}\NormalTok{milliseconds}\OperatorTok{(}\DecValTok{500}\OperatorTok{));}
\NormalTok{    sync\_out}\OperatorTok{.}\NormalTok{println}\OperatorTok{(}\StringTok{"Task "}\OperatorTok{,}\NormalTok{ i}\OperatorTok{,} \StringTok{" done."}\OperatorTok{);}
\OperatorTok{\}}

\DataTypeTok{int}\NormalTok{ main}\OperatorTok{()}
\OperatorTok{\{}
    \ControlFlowTok{for} \OperatorTok{(}\DataTypeTok{size\_t}\NormalTok{ i }\OperatorTok{=} \DecValTok{0}\OperatorTok{;}\NormalTok{ i }\OperatorTok{\textless{}} \DecValTok{8}\OperatorTok{;} \OperatorTok{++}\NormalTok{i}\OperatorTok{)}
\NormalTok{        pool}\OperatorTok{.}\NormalTok{push\_task}\OperatorTok{(}\NormalTok{sleep\_half\_second}\OperatorTok{,}\NormalTok{ i}\OperatorTok{);}
\NormalTok{    sync\_out}\OperatorTok{.}\NormalTok{println}\OperatorTok{(}\StringTok{"Submitted 8 tasks."}\OperatorTok{);}
    \BuiltInTok{std::}\NormalTok{this\_thread}\BuiltInTok{::}\NormalTok{sleep\_for}\OperatorTok{(}\BuiltInTok{std::}\NormalTok{chrono}\BuiltInTok{::}\NormalTok{milliseconds}\OperatorTok{(}\DecValTok{250}\OperatorTok{));}
\NormalTok{    pool}\OperatorTok{.}\NormalTok{paused }\OperatorTok{=} \KeywordTok{true}\OperatorTok{;}
\NormalTok{    sync\_out}\OperatorTok{.}\NormalTok{println}\OperatorTok{(}\StringTok{"Pool paused."}\OperatorTok{);}
    \BuiltInTok{std::}\NormalTok{this\_thread}\BuiltInTok{::}\NormalTok{sleep\_for}\OperatorTok{(}\BuiltInTok{std::}\NormalTok{chrono}\BuiltInTok{::}\NormalTok{milliseconds}\OperatorTok{(}\DecValTok{1000}\OperatorTok{));}
\NormalTok{    sync\_out}\OperatorTok{.}\NormalTok{println}\OperatorTok{(}\StringTok{"Still paused..."}\OperatorTok{);}
    \BuiltInTok{std::}\NormalTok{this\_thread}\BuiltInTok{::}\NormalTok{sleep\_for}\OperatorTok{(}\BuiltInTok{std::}\NormalTok{chrono}\BuiltInTok{::}\NormalTok{milliseconds}\OperatorTok{(}\DecValTok{1000}\OperatorTok{));}
    \ControlFlowTok{for} \OperatorTok{(}\DataTypeTok{size\_t}\NormalTok{ i }\OperatorTok{=} \DecValTok{8}\OperatorTok{;}\NormalTok{ i }\OperatorTok{\textless{}} \DecValTok{12}\OperatorTok{;} \OperatorTok{++}\NormalTok{i}\OperatorTok{)}
\NormalTok{        pool}\OperatorTok{.}\NormalTok{push\_task}\OperatorTok{(}\NormalTok{sleep\_half\_second}\OperatorTok{,}\NormalTok{ i}\OperatorTok{);}
\NormalTok{    sync\_out}\OperatorTok{.}\NormalTok{println}\OperatorTok{(}\StringTok{"Submitted 4 more tasks."}\OperatorTok{);}
\NormalTok{    sync\_out}\OperatorTok{.}\NormalTok{println}\OperatorTok{(}\StringTok{"Still paused..."}\OperatorTok{);}
    \BuiltInTok{std::}\NormalTok{this\_thread}\BuiltInTok{::}\NormalTok{sleep\_for}\OperatorTok{(}\BuiltInTok{std::}\NormalTok{chrono}\BuiltInTok{::}\NormalTok{milliseconds}\OperatorTok{(}\DecValTok{1000}\OperatorTok{));}
\NormalTok{    pool}\OperatorTok{.}\NormalTok{paused }\OperatorTok{=} \KeywordTok{false}\OperatorTok{;}
\NormalTok{    sync\_out}\OperatorTok{.}\NormalTok{println}\OperatorTok{(}\StringTok{"Pool resumed."}\OperatorTok{);}
\OperatorTok{\}}
\end{Highlighting}
\end{Shaded}

Assuming you have at least 4 hardware threads, the output should be
similar to:

\begin{Shaded}
\begin{Highlighting}[]
\NormalTok{Submitted 8 tasks.}
\NormalTok{Pool paused.}
\NormalTok{Task 0 done.}
\NormalTok{Task 1 done.}
\NormalTok{Task 2 done.}
\NormalTok{Task 3 done.}
\NormalTok{Still paused...}
\NormalTok{Submitted 4 more tasks.}
\NormalTok{Still paused...}
\NormalTok{Pool resumed.}
\NormalTok{Task 4 done.}
\NormalTok{Task 5 done.}
\NormalTok{Task 6 done.}
\NormalTok{Task 7 done.}
\NormalTok{Task 8 done.}
\NormalTok{Task 9 done.}
\NormalTok{Task 10 done.}
\NormalTok{Task 11 done.}
\end{Highlighting}
\end{Shaded}

Here is what happened. We initially submitted a total of 8 tasks to the
queue. Since we waited for 250ms before pausing, the first 4 tasks have
already started running, so they kept running until they finished. While
the pool was paused, we submitted 4 more tasks to the queue, but they
just waited at the end of the queue. When we resumed, the remaining 4
initial tasks were executed, followed by the 4 new tasks.

While the workers are paused, \texttt{wait\_for\_tasks()} will wait for
the running tasks instead of all tasks (otherwise it would wait
forever). This is demonstrated by the following program:

\begin{Shaded}
\begin{Highlighting}[]
\PreprocessorTok{\#include }\ImportTok{"BS\_thread\_pool.hpp"}

\NormalTok{BS}\OperatorTok{::}\NormalTok{synced\_stream sync\_out}\OperatorTok{;}
\NormalTok{BS}\OperatorTok{::}\NormalTok{thread\_pool pool}\OperatorTok{(}\DecValTok{4}\OperatorTok{);}

\DataTypeTok{void}\NormalTok{ sleep\_half\_second}\OperatorTok{(}\AttributeTok{const} \DataTypeTok{size\_t}\NormalTok{ i}\OperatorTok{)}
\OperatorTok{\{}
    \BuiltInTok{std::}\NormalTok{this\_thread}\BuiltInTok{::}\NormalTok{sleep\_for}\OperatorTok{(}\BuiltInTok{std::}\NormalTok{chrono}\BuiltInTok{::}\NormalTok{milliseconds}\OperatorTok{(}\DecValTok{500}\OperatorTok{));}
\NormalTok{    sync\_out}\OperatorTok{.}\NormalTok{println}\OperatorTok{(}\StringTok{"Task "}\OperatorTok{,}\NormalTok{ i}\OperatorTok{,} \StringTok{" done."}\OperatorTok{);}
\OperatorTok{\}}

\DataTypeTok{int}\NormalTok{ main}\OperatorTok{()}
\OperatorTok{\{}
    \ControlFlowTok{for} \OperatorTok{(}\DataTypeTok{size\_t}\NormalTok{ i }\OperatorTok{=} \DecValTok{0}\OperatorTok{;}\NormalTok{ i }\OperatorTok{\textless{}} \DecValTok{8}\OperatorTok{;} \OperatorTok{++}\NormalTok{i}\OperatorTok{)}
\NormalTok{        pool}\OperatorTok{.}\NormalTok{push\_task}\OperatorTok{(}\NormalTok{sleep\_half\_second}\OperatorTok{,}\NormalTok{ i}\OperatorTok{);}
\NormalTok{    sync\_out}\OperatorTok{.}\NormalTok{println}\OperatorTok{(}\StringTok{"Submitted 8 tasks. Waiting for them to complete."}\OperatorTok{);}
\NormalTok{    pool}\OperatorTok{.}\NormalTok{wait\_for\_tasks}\OperatorTok{();}
    \ControlFlowTok{for} \OperatorTok{(}\DataTypeTok{size\_t}\NormalTok{ i }\OperatorTok{=} \DecValTok{8}\OperatorTok{;}\NormalTok{ i }\OperatorTok{\textless{}} \DecValTok{20}\OperatorTok{;} \OperatorTok{++}\NormalTok{i}\OperatorTok{)}
\NormalTok{        pool}\OperatorTok{.}\NormalTok{push\_task}\OperatorTok{(}\NormalTok{sleep\_half\_second}\OperatorTok{,}\NormalTok{ i}\OperatorTok{);}
\NormalTok{    sync\_out}\OperatorTok{.}\NormalTok{println}\OperatorTok{(}\StringTok{"Submitted 12 more tasks."}\OperatorTok{);}
    \BuiltInTok{std::}\NormalTok{this\_thread}\BuiltInTok{::}\NormalTok{sleep\_for}\OperatorTok{(}\BuiltInTok{std::}\NormalTok{chrono}\BuiltInTok{::}\NormalTok{milliseconds}\OperatorTok{(}\DecValTok{250}\OperatorTok{));}
\NormalTok{    pool}\OperatorTok{.}\NormalTok{paused }\OperatorTok{=} \KeywordTok{true}\OperatorTok{;}
\NormalTok{    sync\_out}\OperatorTok{.}\NormalTok{println}\OperatorTok{(}\StringTok{"Pool paused. Waiting for the "}\OperatorTok{,}\NormalTok{ pool}\OperatorTok{.}\NormalTok{get\_tasks\_running}\OperatorTok{(),}
                     \StringTok{"running tasks to complete."}\OperatorTok{);}
\NormalTok{    pool}\OperatorTok{.}\NormalTok{wait\_for\_tasks}\OperatorTok{();}
\NormalTok{    sync\_out}\OperatorTok{.}\NormalTok{println}\OperatorTok{(}\StringTok{"All running tasks completed. "}\OperatorTok{,}\NormalTok{ pool}\OperatorTok{.}\NormalTok{get\_tasks\_queued}\OperatorTok{(),}
                     \StringTok{"tasks still queued."}\OperatorTok{);}
    \BuiltInTok{std::}\NormalTok{this\_thread}\BuiltInTok{::}\NormalTok{sleep\_for}\OperatorTok{(}\BuiltInTok{std::}\NormalTok{chrono}\BuiltInTok{::}\NormalTok{milliseconds}\OperatorTok{(}\DecValTok{1000}\OperatorTok{));}
\NormalTok{    sync\_out}\OperatorTok{.}\NormalTok{println}\OperatorTok{(}\StringTok{"Still paused..."}\OperatorTok{);}
    \BuiltInTok{std::}\NormalTok{this\_thread}\BuiltInTok{::}\NormalTok{sleep\_for}\OperatorTok{(}\BuiltInTok{std::}\NormalTok{chrono}\BuiltInTok{::}\NormalTok{milliseconds}\OperatorTok{(}\DecValTok{1000}\OperatorTok{));}
\NormalTok{    sync\_out}\OperatorTok{.}\NormalTok{println}\OperatorTok{(}\StringTok{"Still paused..."}\OperatorTok{);}
    \BuiltInTok{std::}\NormalTok{this\_thread}\BuiltInTok{::}\NormalTok{sleep\_for}\OperatorTok{(}\BuiltInTok{std::}\NormalTok{chrono}\BuiltInTok{::}\NormalTok{milliseconds}\OperatorTok{(}\DecValTok{1000}\OperatorTok{));}
\NormalTok{    pool}\OperatorTok{.}\NormalTok{paused }\OperatorTok{=} \KeywordTok{false}\OperatorTok{;}
    \BuiltInTok{std::}\NormalTok{this\_thread}\BuiltInTok{::}\NormalTok{sleep\_for}\OperatorTok{(}\BuiltInTok{std::}\NormalTok{chrono}\BuiltInTok{::}\NormalTok{milliseconds}\OperatorTok{(}\DecValTok{250}\OperatorTok{));}
\NormalTok{    sync\_out}\OperatorTok{.}\NormalTok{println}\OperatorTok{(}\StringTok{"Pool resumed. Waiting for the remaining "}\OperatorTok{,}\NormalTok{ pool}\OperatorTok{.}\NormalTok{get\_tasks\_total}\OperatorTok{(),}
                     \StringTok{" tasks ("}\OperatorTok{,}\NormalTok{ pool}\OperatorTok{.}\NormalTok{get\_tasks\_running}\OperatorTok{(),} \StringTok{" running and "}\OperatorTok{,}
                     \NormalTok{ pool}\OperatorTok{.}\NormalTok{get\_tasks\_queued}\OperatorTok{(),} \StringTok{" queued) to complete."}\OperatorTok{);}
\NormalTok{    pool}\OperatorTok{.}\NormalTok{wait\_for\_tasks}\OperatorTok{();}
\NormalTok{    sync\_out}\OperatorTok{.}\NormalTok{println}\OperatorTok{(}\StringTok{"All tasks completed."}\OperatorTok{);}
\OperatorTok{\}}
\end{Highlighting}
\end{Shaded}

The output should be similar to:

\begin{Shaded}
\begin{Highlighting}[]
\NormalTok{Submitted 8 tasks. Waiting for them to complete.}
\NormalTok{Task 0 done.}
\NormalTok{Task 1 done.}
\NormalTok{Task 2 done.}
\NormalTok{Task 3 done.}
\NormalTok{Task 4 done.}
\NormalTok{Task 5 done.}
\NormalTok{Task 6 done.}
\NormalTok{Task 7 done.}
\NormalTok{Submitted 12 more tasks.}
\NormalTok{Pool paused. Waiting for the 4 running tasks to complete.}
\NormalTok{Task 8 done.}
\NormalTok{Task 9 done.}
\NormalTok{Task 10 done.}
\NormalTok{Task 11 done.}
\NormalTok{All running tasks completed. 8 tasks still queued.}
\NormalTok{Still paused...}
\NormalTok{Still paused...}
\NormalTok{Pool resumed. Waiting for the remaining 8 tasks (4 running and 4 queued) to complete.}
\NormalTok{Task 12 done.}
\NormalTok{Task 13 done.}
\NormalTok{Task 14 done.}
\NormalTok{Task 15 done.}
\NormalTok{Task 16 done.}
\NormalTok{Task 17 done.}
\NormalTok{Task 18 done.}
\NormalTok{Task 19 done.}
\NormalTok{All tasks completed.}
\end{Highlighting}
\end{Shaded}

The first \texttt{wait\_for\_tasks()}, which was called with
\texttt{paused\ ==\ false}, waited for all 8 tasks, both running and
queued. The second \texttt{wait\_for\_tasks()}, which was called with
\texttt{paused\ ==\ true}, only waited for the 4 running tasks, while
the other 8 tasks remained queued, and were not executed since the pool
was paused. Finally, the third \texttt{wait\_for\_tasks()}, which was
called with \texttt{paused\ ==\ false}, waited for the remaining 8
tasks, both running and queued.

\textbf{Warning}: If the thread pool is destroyed while paused, any
tasks still in the queue will never be executed!

\hypertarget{exception-handling}{%
\subsubsection{Exception handling}\label{exception-handling}}

\texttt{submit()} catches any exceptions thrown by the submitted task
and forwards them to the corresponding future. They can then be caught
when invoking the \texttt{get()} member function of the future. For
example:

\begin{Shaded}
\begin{Highlighting}[]
\PreprocessorTok{\#include }\ImportTok{"BS\_thread\_pool.hpp"}

\DataTypeTok{double}\NormalTok{ inverse}\OperatorTok{(}\AttributeTok{const} \DataTypeTok{double}\NormalTok{ x}\OperatorTok{)}
\OperatorTok{\{}
    \ControlFlowTok{if} \OperatorTok{(}\NormalTok{x }\OperatorTok{==} \DecValTok{0}\OperatorTok{)}
        \ControlFlowTok{throw} \BuiltInTok{std::}\NormalTok{runtime\_error}\OperatorTok{(}\StringTok{"Division by zero!"}\OperatorTok{);}
    \ControlFlowTok{else}
        \ControlFlowTok{return} \DecValTok{1} \OperatorTok{/}\NormalTok{ x}\OperatorTok{;}
\OperatorTok{\}}

\DataTypeTok{int}\NormalTok{ main}\OperatorTok{()}
\OperatorTok{\{}
\NormalTok{    BS}\OperatorTok{::}\NormalTok{thread\_pool pool}\OperatorTok{;}
    \KeywordTok{auto}\NormalTok{ my\_future }\OperatorTok{=}\NormalTok{ pool}\OperatorTok{.}\NormalTok{submit}\OperatorTok{(}\NormalTok{inverse}\OperatorTok{,} \DecValTok{0}\OperatorTok{);}
    \ControlFlowTok{try}
    \OperatorTok{\{}
        \AttributeTok{const} \DataTypeTok{double}\NormalTok{ result }\OperatorTok{=}\NormalTok{ my\_future}\OperatorTok{.}\NormalTok{get}\OperatorTok{();}
        \BuiltInTok{std::}\NormalTok{cout}\OperatorTok{ \textless{}\textless{}} \StringTok{"The result is: "} \OperatorTok{\textless{}\textless{}}\NormalTok{ result }\OperatorTok{\textless{}\textless{}} \CharTok{\textquotesingle{}}\SpecialCharTok{\textbackslash{}n}\CharTok{\textquotesingle{}}\OperatorTok{;}
    \OperatorTok{\}}
    \ControlFlowTok{catch} \OperatorTok{(}\AttributeTok{const} \BuiltInTok{std::}\NormalTok{exception}\OperatorTok{\&}\NormalTok{ e}\OperatorTok{)}
    \OperatorTok{\{}
        \BuiltInTok{std::}\NormalTok{cout}\OperatorTok{ \textless{}\textless{}} \StringTok{"Caught exception: "} \OperatorTok{\textless{}\textless{}}\NormalTok{ e}\OperatorTok{.}\NormalTok{what}\OperatorTok{()} \OperatorTok{\textless{}\textless{}} \CharTok{\textquotesingle{}}\SpecialCharTok{\textbackslash{}n}\CharTok{\textquotesingle{}}\OperatorTok{;}
    \OperatorTok{\}}
\OperatorTok{\}}
\end{Highlighting}
\end{Shaded}

The output will be:

\begin{Shaded}
\begin{Highlighting}[]
\NormalTok{Caught exception: Division by zero!}
\end{Highlighting}
\end{Shaded}

\hypertarget{testing-the-package}{%
\subsection{Testing the package}\label{testing-the-package}}

The included file \texttt{BS\_thread\_pool\_test.cpp} will perform
automated tests of all aspects of the package, and perform simple
benchmarks. The output will be printed both to \texttt{std::cout} and to
a file named \texttt{BS\_thread\_pool\_test-yyyy-mm-dd\_hh.mm.ss.log}
based on the current date and time. In addition, the code is thoroughly
documented, and is meant to serve as an extensive example of how to
properly use the package.

Please make sure to:

\begin{enumerate}
\tightlist
\item
  \protect\hyperlink{compiling-and-compatibility}{Compile}
  \texttt{BS\_thread\_pool\_test.cpp} with optimization flags enabled
  (e.g. \texttt{-O3} on GCC / Clang or \texttt{/O2} on MSVC).
\item
  Run the test without any other applications, especially multithreaded
  applications, running in parallel.
\end{enumerate}

If any of the tests fail, please
\href{https://github.com/bshoshany/thread-pool/issues}{submit a bug
report} including the exact specifications of your system (OS, CPU,
compiler, etc.) and the generated log file.

\hypertarget{automated-tests}{%
\subsubsection{Automated tests}\label{automated-tests}}

A sample output of a successful run of the automated tests is as
follows:

\begin{Shaded}
\begin{Highlighting}[]
\NormalTok{A C++17 Thread Pool for High{-}Performance Scientific Computing}
\NormalTok{(c) 2022 Barak Shoshany (baraksh@gmail.com) (http://baraksh.com)}
\NormalTok{GitHub: https://github.com/bshoshany/thread{-}pool}

\NormalTok{Thread pool library version is v3.0.0 (2022{-}05{-}30).}
\NormalTok{Hardware concurrency is 24.}
\NormalTok{Generating log file: BS\_thread\_pool\_test{-}2022{-}05{-}30\_22.59.30.log.}

\NormalTok{Important: Please do not run any other applications, especially multithreaded}
\NormalTok{applications, in parallel with this test!}

\NormalTok{====================================}
\NormalTok{Checking that the constructor works:}
\NormalTok{====================================}
\NormalTok{Checking that the thread pool reports a number of threads equal to the hardware}
\NormalTok{concurrency...}
\NormalTok{{-}\textgreater{} PASSED!}
\NormalTok{Checking that the manually counted number of unique thread IDs is equal to the}
\NormalTok{reported number of threads...}
\NormalTok{{-}\textgreater{} PASSED!}

\NormalTok{============================}
\NormalTok{Checking that reset() works:}
\NormalTok{============================}
\NormalTok{Checking that after reset() the thread pool reports a number of threads equal to}
\NormalTok{half the hardware concurrency...}
\NormalTok{{-}\textgreater{} PASSED!}
\NormalTok{Checking that after reset() the manually counted number of unique thread IDs is}
\NormalTok{equal to the reported number of threads...}
\NormalTok{{-}\textgreater{} PASSED!}
\NormalTok{Checking that after a second reset() the thread pool reports a number of threads}
\NormalTok{equal to the hardware concurrency...}
\NormalTok{{-}\textgreater{} PASSED!}
\NormalTok{Checking that after a second reset() the manually counted number of unique thread}
\NormalTok{IDs is equal to the reported number of threads...}
\NormalTok{{-}\textgreater{} PASSED!}

\NormalTok{================================}
\NormalTok{Checking that push\_task() works:}
\NormalTok{================================}
\NormalTok{Checking that push\_task() works for a function with no arguments or return value...}
\NormalTok{{-}\textgreater{} PASSED!}
\NormalTok{Checking that push\_task() works for a function with one argument and no return value...}
\NormalTok{{-}\textgreater{} PASSED!}
\NormalTok{Checking that push\_task() works for a function with two arguments and no return value...}
\NormalTok{{-}\textgreater{} PASSED!}

\NormalTok{=============================}
\NormalTok{Checking that submit() works:}
\NormalTok{=============================}
\NormalTok{Checking that submit() works for a function with no arguments or return value...}
\NormalTok{{-}\textgreater{} PASSED!}
\NormalTok{Checking that submit() works for a function with one argument and no return value...}
\NormalTok{{-}\textgreater{} PASSED!}
\NormalTok{Checking that submit() works for a function with two arguments and no return value...}
\NormalTok{{-}\textgreater{} PASSED!}
\NormalTok{Checking that submit() works for a function with no arguments and a return value...}
\NormalTok{{-}\textgreater{} PASSED!}
\NormalTok{Checking that submit() works for a function with one argument and a return value...}
\NormalTok{{-}\textgreater{} PASSED!}
\NormalTok{Checking that submit() works for a function with two arguments and a return value...}
\NormalTok{{-}\textgreater{} PASSED!}

\NormalTok{=======================================}
\NormalTok{Checking that wait\_for\_tasks() works...}
\NormalTok{=======================================}
\NormalTok{{-}\textgreater{} PASSED!}

\NormalTok{=======================================}
\NormalTok{Checking that parallelize\_loop() works:}
\NormalTok{=======================================}
\NormalTok{Verifying that a loop from {-}434827 to 461429 with 23 tasks modifies all indices...}
\NormalTok{{-}\textgreater{} PASSED!}
\NormalTok{Verifying that a loop from 255333 to {-}889028 with 9 tasks modifies all indices...}
\NormalTok{{-}\textgreater{} PASSED!}
\NormalTok{Verifying that a loop from {-}257322 to 550471 with 5 tasks modifies all indices...}
\NormalTok{{-}\textgreater{} PASSED!}
\NormalTok{Verifying that a loop from {-}257648 to {-}475958 with 23 tasks modifies all indices...}
\NormalTok{{-}\textgreater{} PASSED!}
\NormalTok{Verifying that a loop from 175412 to {-}544672 with 13 tasks modifies all indices...}
\NormalTok{{-}\textgreater{} PASSED!}
\NormalTok{Verifying that a loop from {-}244797 to {-}970178 with 11 tasks modifies all indices...}
\NormalTok{{-}\textgreater{} PASSED!}
\NormalTok{Verifying that a loop from 411251 to {-}718341 with 15 tasks modifies all indices...}
\NormalTok{{-}\textgreater{} PASSED!}
\NormalTok{Verifying that a loop from 418787 to 978302 with 22 tasks modifies all indices...}
\NormalTok{{-}\textgreater{} PASSED!}
\NormalTok{Verifying that a loop from {-}2412 to {-}310158 with 4 tasks modifies all indices...}
\NormalTok{{-}\textgreater{} PASSED!}
\NormalTok{Verifying that a loop from {-}881862 to 137673 with 7 tasks modifies all indices...}
\NormalTok{{-}\textgreater{} PASSED!}
\NormalTok{Verifying that a loop from 539438 to {-}759983 with 17 tasks correctly sums all indices...}
\NormalTok{{-}\textgreater{} PASSED!}
\NormalTok{Verifying that a loop from 745706 to {-}519554 with 13 tasks correctly sums all indices...}
\NormalTok{{-}\textgreater{} PASSED!}
\NormalTok{Verifying that a loop from 288078 to 432534 with 12 tasks correctly sums all indices...}
\NormalTok{{-}\textgreater{} PASSED!}
\NormalTok{Verifying that a loop from {-}487251 to 302796 with 4 tasks correctly sums all indices...}
\NormalTok{{-}\textgreater{} PASSED!}
\NormalTok{Verifying that a loop from 408766 to 756890 with 3 tasks correctly sums all indices...}
\NormalTok{{-}\textgreater{} PASSED!}
\NormalTok{Verifying that a loop from {-}976768 to {-}500744 with 10 tasks correctly sums all indices...}
\NormalTok{{-}\textgreater{} PASSED!}
\NormalTok{Verifying that a loop from {-}817442 to 175967 with 6 tasks correctly sums all indices...}
\NormalTok{{-}\textgreater{} PASSED!}
\NormalTok{Verifying that a loop from {-}765007 to {-}53682 with 23 tasks correctly sums all indices...}
\NormalTok{{-}\textgreater{} PASSED!}
\NormalTok{Verifying that a loop from {-}903190 to {-}361760 with 14 tasks correctly sums all indices...}
\NormalTok{{-}\textgreater{} PASSED!}
\NormalTok{Verifying that a loop from 72823 to {-}85485 with 15 tasks correctly sums all indices...}
\NormalTok{{-}\textgreater{} PASSED!}

\NormalTok{====================================}
\NormalTok{Checking that task monitoring works:}
\NormalTok{====================================}
\NormalTok{Resetting pool to 4 threads.}
\NormalTok{Submitting 12 tasks.}
\NormalTok{After submission, should have:}
\NormalTok{12 tasks total, 4 tasks running, 8 tasks queued...}
\NormalTok{{-}\textgreater{} PASSED!}
\NormalTok{Task 0 released.}
\NormalTok{Task 2 released.}
\NormalTok{Task 1 released.}
\NormalTok{Task 3 released.}
\NormalTok{After releasing 4 tasks, should have:}
\NormalTok{8 tasks total, 4 tasks running, 4 tasks queued...}
\NormalTok{{-}\textgreater{} PASSED!}
\NormalTok{Task 7 released.}
\NormalTok{Task 5 released.}
\NormalTok{Task 4 released.}
\NormalTok{Task 6 released.}
\NormalTok{After releasing 4 more tasks, should have:}
\NormalTok{4 tasks total, 4 tasks running, 0 tasks queued...}
\NormalTok{{-}\textgreater{} PASSED!}
\NormalTok{Task 10 released.}
\NormalTok{Task 8 released.}
\NormalTok{Task 9 released.}
\NormalTok{Task 11 released.}
\NormalTok{After releasing the final 4 tasks, should have:}
\NormalTok{0 tasks total, 0 tasks running, 0 tasks queued...}
\NormalTok{{-}\textgreater{} PASSED!}
\NormalTok{Resetting pool to 24 threads.}

\NormalTok{============================}
\NormalTok{Checking that pausing works:}
\NormalTok{============================}
\NormalTok{Resetting pool to 4 threads.}
\NormalTok{Pausing pool.}
\NormalTok{Submitting 12 tasks, each one waiting for 200ms.}
\NormalTok{Immediately after submission, should have:}
\NormalTok{12 tasks total, 0 tasks running, 12 tasks queued...}
\NormalTok{{-}\textgreater{} PASSED!}
\NormalTok{300ms later, should still have:}
\NormalTok{12 tasks total, 0 tasks running, 12 tasks queued...}
\NormalTok{{-}\textgreater{} PASSED!}
\NormalTok{Unpausing pool.}
\NormalTok{Task 1 done.}
\NormalTok{Task 2 done.}
\NormalTok{Task 3 done.}
\NormalTok{Task 0 done.}
\NormalTok{300ms later, should have:}
\NormalTok{8 tasks total, 4 tasks running, 4 tasks queued...}
\NormalTok{{-}\textgreater{} PASSED!}
\NormalTok{Pausing pool and using wait\_for\_tasks() to wait for the running tasks.}
\NormalTok{Task 6 done.}
\NormalTok{Task 4 done.}
\NormalTok{Task 5 done.}
\NormalTok{Task 7 done.}
\NormalTok{After waiting, should have:}
\NormalTok{4 tasks total, 0 tasks running, 4 tasks queued...}
\NormalTok{{-}\textgreater{} PASSED!}
\NormalTok{200ms later, should still have:}
\NormalTok{4 tasks total, 0 tasks running, 4 tasks queued...}
\NormalTok{{-}\textgreater{} PASSED!}
\NormalTok{Unpausing pool and using wait\_for\_tasks() to wait for all tasks.}
\NormalTok{Task 9 done.}
\NormalTok{Task 10 done.}
\NormalTok{Task 11 done.}
\NormalTok{Task 8 done.}
\NormalTok{After waiting, should have:}
\NormalTok{0 tasks total, 0 tasks running, 0 tasks queued...}
\NormalTok{{-}\textgreater{} PASSED!}
\NormalTok{Resetting pool to 24 threads.}

\NormalTok{=======================================}
\NormalTok{Checking that exception handling works:}
\NormalTok{=======================================}
\NormalTok{{-}\textgreater{} PASSED!}

\NormalTok{============================================================}
\NormalTok{Testing that vector operations produce the expected results:}
\NormalTok{============================================================}
\NormalTok{Adding two vectors with 83788 elements using 24 tasks...}
\NormalTok{{-}\textgreater{} PASSED!}
\NormalTok{Adding two vectors with 595750 elements using 3 tasks...}
\NormalTok{{-}\textgreater{} PASSED!}
\NormalTok{Adding two vectors with 738336 elements using 20 tasks...}
\NormalTok{{-}\textgreater{} PASSED!}
\NormalTok{Adding two vectors with 100123 elements using 24 tasks...}
\NormalTok{{-}\textgreater{} PASSED!}
\NormalTok{Adding two vectors with 921883 elements using 24 tasks...}
\NormalTok{{-}\textgreater{} PASSED!}
\NormalTok{Adding two vectors with 76713 elements using 22 tasks...}
\NormalTok{{-}\textgreater{} PASSED!}
\NormalTok{Adding two vectors with 891037 elements using 2 tasks...}
\NormalTok{{-}\textgreater{} PASSED!}
\NormalTok{Adding two vectors with 245369 elements using 17 tasks...}
\NormalTok{{-}\textgreater{} PASSED!}
\NormalTok{Adding two vectors with 39624 elements using 11 tasks...}
\NormalTok{{-}\textgreater{} PASSED!}
\NormalTok{Adding two vectors with 295307 elements using 10 tasks...}
\NormalTok{{-}\textgreater{} PASSED!}

\NormalTok{++++++++++++++++++++++++++++++}
\NormalTok{SUCCESS: Passed all 57 checks!}
\NormalTok{++++++++++++++++++++++++++++++}
\end{Highlighting}
\end{Shaded}

\hypertarget{performance-tests}{%
\subsubsection{Performance tests}\label{performance-tests}}

If all checks passed, \texttt{BS\_thread\_pool\_test.cpp} will perform
simple benchmarks by filling a specific number of vectors of fixed size
with random values. The program decides how many vectors to use by
testing how many are needed to reach a target duration in the
single-threaded test. This ensures that the test takes approximately the
same amount of time on different systems, and is thus more consistent
and portable.

Once the required number of vectors has been determined, the program
will test the performance of several multi-threaded tests, dividing the
total number of vectors into different numbers of tasks, compare them to
the performance of the single-threaded test, and indicate the maximum
speedup obtained.

Please note that these benchmarks are only intended to demonstrate that
the package can provide a significant speedup, and it is highly
recommended to perform your own benchmarks with your specific system,
compiler, and code.

Here we will present the results of the performance test running on a
high-end desktop computer equipped with a 12-core / 24-thread AMD Ryzen
9 3900X CPU at 3.8 GHz and 32 GB of DDR4 RAM at 3600 MHz, compiled using
\href{https://gcc.gnu.org/}{GCC} v12.1.0
(\href{https://winlibs.com/}{WinLibs build}) on Windows 11 build
22000.675 with the \texttt{-O3} compiler flag. The output was as
follows:

\begin{Shaded}
\begin{Highlighting}[]
\NormalTok{======================}
\NormalTok{Performing benchmarks:}
\NormalTok{======================}
\NormalTok{Using 24 threads.}
\NormalTok{Each test will be repeated 20 times to collect reliable statistics.}

\NormalTok{Generating 57320 random vectors with 500 elements each:}
\NormalTok{Single{-}threaded, mean execution time was  298.2 ms with standard deviation  1.7 ms.}
\NormalTok{With    6 tasks, mean execution time was   52.3 ms with standard deviation  1.3 ms.}
\NormalTok{With   12 tasks, mean execution time was   30.3 ms with standard deviation  0.8 ms.}
\NormalTok{With   24 tasks, mean execution time was   16.4 ms with standard deviation  1.2 ms.}
\NormalTok{With   48 tasks, mean execution time was   19.2 ms with standard deviation  2.6 ms.}
\NormalTok{With   96 tasks, mean execution time was   17.8 ms with standard deviation  1.2 ms.}
\NormalTok{Maximum speedup obtained by multithreading vs. single{-}threading: 18.2x, using 24 tasks.}

\NormalTok{+++++++++++++++++++++++++++++++++++++++}
\NormalTok{Thread pool performance test completed!}
\NormalTok{+++++++++++++++++++++++++++++++++++++++}
\end{Highlighting}
\end{Shaded}

This CPU has 12 physical cores, with each core providing two separate
logical cores via hyperthreading, for a total of 24 threads. Without
hyperthreading, we would expect a maximum theoretical speedup of 12x.
With hyperthreading, one might naively expect to achieve up to a 24x
speedup, but this is in fact impossible, as both logical cores share the
same physical core's resources. However, generally we would expect an
estimated 30\% additional speedup \cite{Casey2011} from hyperthreading, which amounts to
around 15.6x in this case. In our performance test, we see a speedup of
18.2x, saturating and even surpassing this estimated theoretical upper
bound.

\bibliographystyle{ieeetr}
\bibliography{thread_pool}

\begin{thebibliography}{1}

\bibitem{Williams2019}
A.~Williams, {\em C++ Concurrency in Action}.
\newblock Manning Publications, 2019.

\bibitem{Stroustrup2013}
B.~Stroustrup, {\em The C++ Programming Language}.
\newblock Pearson Education, 2013.

\bibitem{Stroustrup2014}
B.~Stroustrup, {\em Programming: Principles and Practice Using C++}.
\newblock Pearson Education, 2014.

\bibitem{Stroustrup2018}
B.~Stroustrup, {\em A Tour of C++}.
\newblock C++ In-Depth Series, Pearson Education, 2018.

\bibitem{ISO2017}
I.~14882:2017, {\em Programming languages - C++}.
\newblock ISO, 2017.

\bibitem{Casey2011}
S.~D. Casey, ``How to determine the effectiveness of hyper-threading technology
  with an application,'' {\em Intel Technology Journal}, vol.~6, no.~1 p.11,
  2011.

\end{thebibliography}

\end{document}